\documentclass[english,prc,twocolumn]{revtex4-2}
\usepackage[T1]{fontenc}
\usepackage[latin9]{inputenc}
\usepackage{xcolor}
\usepackage{babel}
\usepackage{float}
\usepackage{amsmath}
\usepackage{amssymb}
\usepackage{graphicx}
\usepackage[unicode=true,pdfusetitle,
 bookmarks=true,bookmarksnumbered=false,bookmarksopen=false,
 breaklinks=false,pdfborder={0 0 1},backref=false,colorlinks=true]
 {hyperref}

\usepackage{longtable}
\usepackage{multirow}
\usepackage{booktabs}

\newcommand{\hl}[1]{\textcolor{black}{#1}}
\newcommand{\bz}[1]{#1}



\begin{document}
\title{The domain of soft hadrons in transverse momentum space in high energy collisions}
\author{Jun Song }
\affiliation{School of Physics and Electronic Engineering, Jining University, Shandong
273155, China}
\author{Hai-hong Li}
\affiliation{School of Physics and Electronic Engineering, Jining University, Shandong
273155, China}
\author{Feng-lan Shao}
\affiliation{School of Physics and Physical Engineering, Qufu Normal University,
Shandong 273165, China}
\begin{abstract}
    By studying experimental data for invariant transverse momentum distribution $f(p_{T})$ of hadrons in high energy $pp$, $p$A and AA collisions, we find two characteristic $p_{T}$ points relating to the behavior of $\left[\ln f(p_{T})\right]^{''}$, i.e., the second derivative of the logarithm of hadronic $p_{T}$ spectrum with respect to $p_{T}$.  One point is $p_{0}$ at which $\left[\ln f(p_{T})\right]^{''}$ is zero and another point is $p_{1}$ at which $\left[\ln f(p_{T})\right]^{''}$ reaches maximum. \bz{The hadronic distribution can be classified into three kinetic regions, i.e., soft region $0<p_{T}<p_{0}$ dominated by hadrons from soft parton system, hard region $p_{T}>p_{1}$ dominated by hadrons from high energy partonic jet, and the transition region $p_{0}<p_{T}<p_{1}$ of above two sources of hadron production.} Using rich data of hadronic $p_{T}$ spectra at RHIC and LHC, we carry out a systematical analysis for $p_{0}$ and $p_{1}$ of kaon, $K^{*}$, $\phi$ and baryons such as $p$, $\Lambda$, $\Xi$, $\Omega$ in $pp$, $p$-Pb and heavy-ion collisions, and show their dependence on hadron species, collision energy, collision centrality and/or charged-particle multiplicity. We also study the correlation between $p_{0}$ of hadrons and their average transverse momentum $\left\langle p_{T}\right\rangle $, and find a difference between $p_{0}$-$\left\langle p_{T}\right\rangle $ correlation of baryons and that of mesons which can be understood by quark combination mechanism at hadronization. The systematic comparison for $p_{0}$-$\left\langle p_{T}\right\rangle $ correlation of kaon, $\phi$, $p$ and $\Lambda$ in $pp$, $p$-Pb, Pb-Pb and Au+Au collisions at RHIC and LHC energies indicates that $p_{0}$ is a sensitive physical quantity of reflecting the production property of soft hadrons in high energy collisions. 
\end{abstract}
\maketitle

\section{introduction}

Collisions of $pp$, $p$A and AA at extremely relativistic energies
are complex processes of producing multiple particles \citep{Wong:1995jf,Bertulani:2002kfo}.
Elementary collisions of partons from colliding particles/nuclei are generally classified into soft processes and hard processes according to the
momentum transfer in the partonic processes. Hard process has the
feature of large momentum transfer and can be calculated by perturbative
QCD method \citep{Feynman:1978dt}. Final-state outcomes of hard processes
are various hadrons usually with large transverse momentum ($p_{T}$)
in the given longitudinal rapidity window. Besides the relatively
rare hard QCD processes, there are lots of soft QCD processes in collisions
and their final-state outcomes are usually hadrons with low $p_{T}$.
Final-state hadrons measured in experiments are composite results
of above two processes and their production properties are usually
described by the differential distributions with respect to $p_{T}$,
rapidity as well as other kinetic variables.

The $p_{T}$ spectrum of the produced hadrons is an important observable
in high energy collisions. In general, hadronic $p_{T}$ spectrum
in the low $p_{T}$ range is different in shape from that in high
$p_{T}$ range to a certain extent and the change is smoothly happened
in the transition region. Although the concept that the production
of hadrons with low/large $p_{T}$ is dominated by soft/hard QCD
process is a general consensus, it is practically unclear about the
precise boundary of the low $p_{T}$ range (i.e., up limit of the
range since low limit is zero) in study of soft hadrons, because the
description of soft QCD process in high energy collisions is not
fully solved from the rigorous non-perturbative QCD calculation. This
uncertainty further leads to the imprecision of the transition range
from soft to hard processes and also the imprecision for the low limit
of the large $p_{T}$ range dominated by hard hadrons.

$p_{T}$ spectra of identified hadrons are widely measured in (at)
different collision systems (energies) from SPS in early years \citep{NA49:2002pzu,NA49:2004jzr,NA49:2006gaj,NA49:2007stj}
to RHIC and LHC experiments in recent years \citep{STAR:2003jwm,STAR:2008med,STAR:2011iap,STAR:2017sal,ALICE:2013mez,ALICE:2013wgn}.
Experimental experts usually use Boltzmann-like distributions to fit
data of hadronic $p_{T}$ spectra in the low $p_{T}$ range to obtain
the effective excitation temperature \citep{NA49:2006gaj,NA49:2007stj}
or use a hydrodynamic-motivated blast-wave model to obtain the radial
flow and kinetic freeze-out temperature \citep{STAR:2003jwm,STAR:2008med,STAR:2017sal,ALICE:2013mez,ALICE:2013wgn}.
$p_{T}$ spectra of hadrons at large $p_{T}$ are usually compared
with various theoretical calculations based on perturbative QCD and
fragmentation \citep{deFlorian:2017lwf,Albino:2008fy,STAR:2011iap,deFlorian:2007ekg},
which usually exhibit a power-law behavior at large $p_{T}$. Theoretical
models incorporating hydrodynamics and jet physics can model the interplay
between soft and hard hadronic components in $p_{T}$ distribution
\citep{Hirano:2003pw,Eskola:2005ue}. In view of these phenomenological
fits, theoretical calculations and, in particular, the precise data
of hadronic $p_{T}$ spectra, we therefore consider whether some regularities
exist in experimental data of $p_{T}$ spectra of hadrons, which can
be used to constrain the kinetic ranges of soft hadrons and
hard hadrons as well as their interplay region. 

\begin{figure}[H]
\begin{centering}
\includegraphics[scale=0.4]{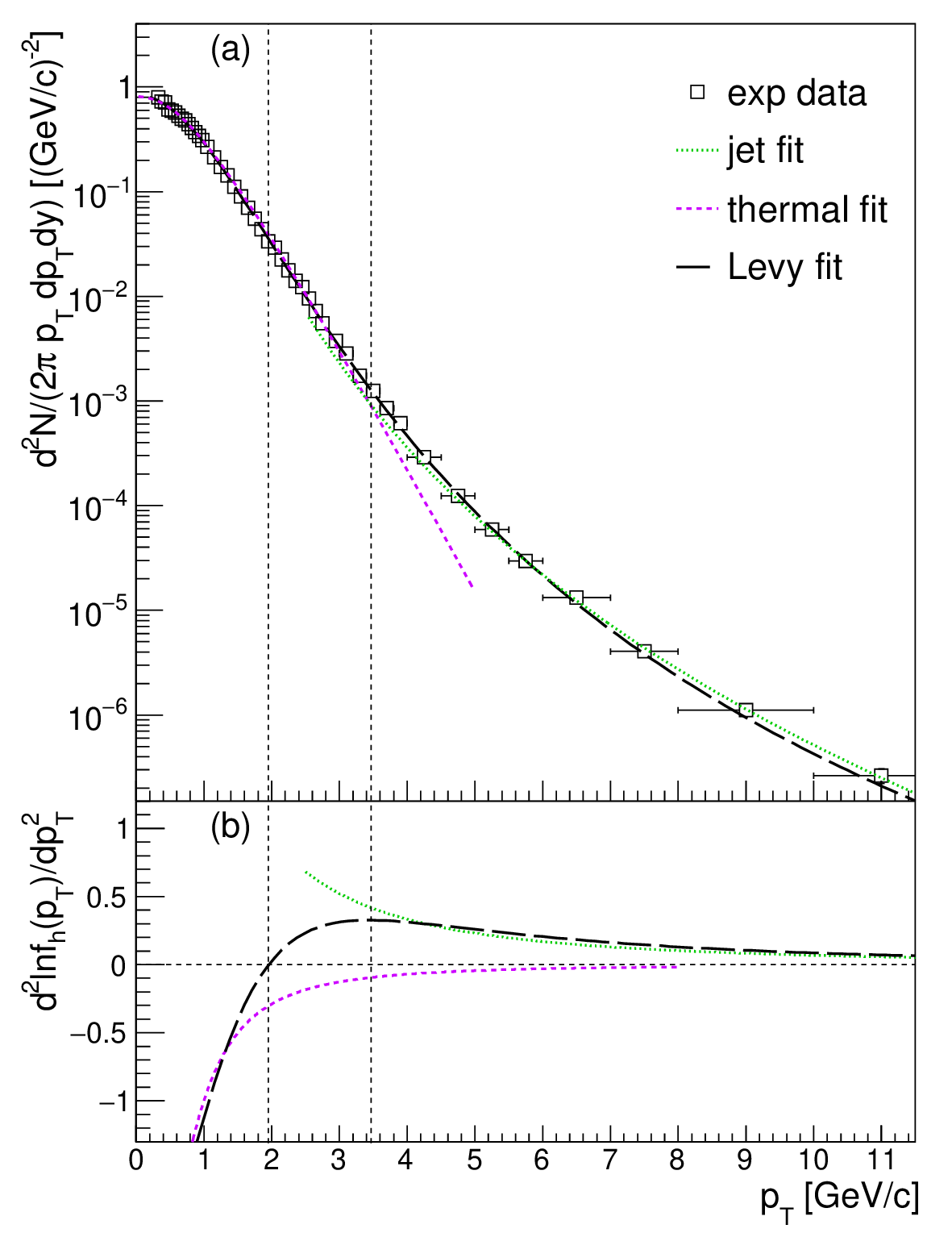}
\par\end{centering}
\caption{(a) $p_{T}$ spectrum of proton at mid-rapidity in 60-80\% centrality in Pb-Pb collisions at $\sqrt{s_{NN}}=2.76$ TeV \citep{ALICE:2014juv} and three fitting lines; (b) the second derivative of the logarithm of the fitting functions with respect to $p_{T}$. \label{fig:p_c6080}}
\end{figure}

After carefully and systematically analyzing $p_{T}$ spectra of identified
hadrons in high energy collisions, we find a probable regularity for
$p_{T}$ spectra of hadrons. Taking $p_{T}$ spectrum of proton at
mid-rapidity in peripheral Pb-Pb collisions at $\sqrt{s_{NN}}=$ 2.76
TeV as the first example, as shown in Fig.~\ref{fig:p_c6080}, we
examine the property of invariant distribution $f_{h}\left(p_{T}\right)\equiv dN^2/(2\pi p_{T}dp_{T}dy)$
of proton in the logarithmic view on vertical axis and find that the
curvature of the distribution exhibits a non-monotonic $p_{T}$ dependence
and changes sign at a certain $p_{T}$. To demonstrate it, we apply
the Levy-Tsallis function \citep{Tsallis:1987eu} 

\begin{equation}
\frac{dN^{Levy}}{2\pi p_{T}dp_{T}}=Np_{T}\left[1+\frac{m_{T}-m}{nc}\right]^{-n}\label{eq:fpt_levy_v0}
\end{equation}
with $m_{T}=\sqrt{p_{T}^{2}+m^{2}}$ to fit the spectrum in a broad
$p_{T}$ range. Here, $N$ is normalization coefficient and $n,c$
are parameters. \hl{$m$ is treated as free parameter in order to make the best fit.}
The long dashed line in top panel in Fig.~\ref{fig:p_c6080} is the fitting result.
In order to focus on the low $p_{T}$ range, we also use a thermal
distribution boosted with a transverse collective velocity $v$ (in
units of $c$)
\begin{equation}
\frac{dN^{th}}{2\pi p_{T}dp_{T}}=Nm_{T}\exp\left[-\frac{\gamma\left(m_{T}-p_{T}v\right)}{T}\right]\label{eq:fpt_thermal}
\end{equation}
with $\gamma=\sqrt{1-v^{2}}$ to fit experimental data at low $p_{T}$.
Here, $N,T,m,v$ are fit parameters. The dashed line in top panel
is the fitting result. In order to focus on large $p_{T}$, we use
a jet-like function 
\begin{equation}
\frac{dN^{jet}}{2\pi p_{T}dp_{T}}=A\left(1+\frac{p_{T}}{b}\right)^{-n}\label{eq:fpt_jet}
\end{equation}
to fit data at large $p_{T}$. $A,b,n$ are fit parameters. The dotted
line in top panel is the fitting result. From the top panel in Fig.~\ref{fig:p_c6080}, we see that thermal function in Eq.~(\ref{eq:fpt_thermal})
can well fit the data in the low $p_{T}$ range and jet function in
Eq.~(\ref{eq:fpt_jet}) can well fit data at large $p_{T}$. Levy-Tsallis
function in Eq.~(\ref{eq:fpt_levy_v0}) can provide a good fit in
both low and large $p_{T}$ range. 

Now, we consider the second derivative of the logarithm of the spectrum
with respect to $p_{T}$
\begin{equation}
\left[\ln f_{h}\left(p_{T}\right)\right]^{''}\equiv\frac{d^{2}\ln f_{h}\left(p_{T}\right)}{dp_{T}^{2}}\label{eq:lnfpt_p2_def}
\end{equation}
for three fitting functions and results are shown in bottom panel
of Fig.~\ref{fig:p_c6080}. We see that the curvature of the thermal
distribution is always negative, i.e., $\left[\ln f_{h}^{th}\left(p_{T}\right)\right]^{''}<0$,
and tends to zero at large $p_{T}$. The curvature of the jet distribution
is always positive, i.e., $\left[\ln f_{h}^{jet}\left(p_{T}\right)\right]^{''}>0$,
and tends to zero at very large $p_{T}$. The curvature of the Levy-Tsallis
function well reproduces thermal distribution at low $p_{T}$ where
$\left[\ln f_{h}^{Levy}\left(p_{T}\right)\right]^{''}<0$ and jet
distribution at high $p_{T}$ where $\left[\ln f_{h}^{Levy}\left(p_{T}\right)\right]^{''}>0$
and properly connects the both in the intermediate $p_{T}$ range where
$\left[\ln f_{h}^{Levy}\left(p_{T}\right)\right]^{''}$changes the
sign, peaks and turns to decrease with $p_{T}$. 

From the non-monotonic behavior of Levy-Tsallis fit, we obtain two
physical points, i.e., the first point $p_{0}\approx2.0$ GeV/$c$ where
$\left[\ln f_{h}\left(p_{T}\right)\right]^{''}$ reaches zero and
the second point $p_{1}\approx3.5$ GeV/$c$ where $\left[\ln f_{h}\left(p_{T}\right)\right]^{''}$
reaches the maximum value. \bz{Following two guiding vertical lines crossing two panels, we see that the $p_{T}$ spectrum of proton can be split into three kinetic regions. Region $0<p_{T}<p_{0}$ denotes the area dominated by thermal or soft hadrons, and region $p_{T}>p_{1}$ denotes the area dominated by hard hadrons from high energy partons or jets. Region $p_{0}<p_{T}<p_{1}$ denotes the area of interplay between soft and hard sources. }

\begin{figure}[H]
\begin{centering}
\includegraphics[scale=0.4]{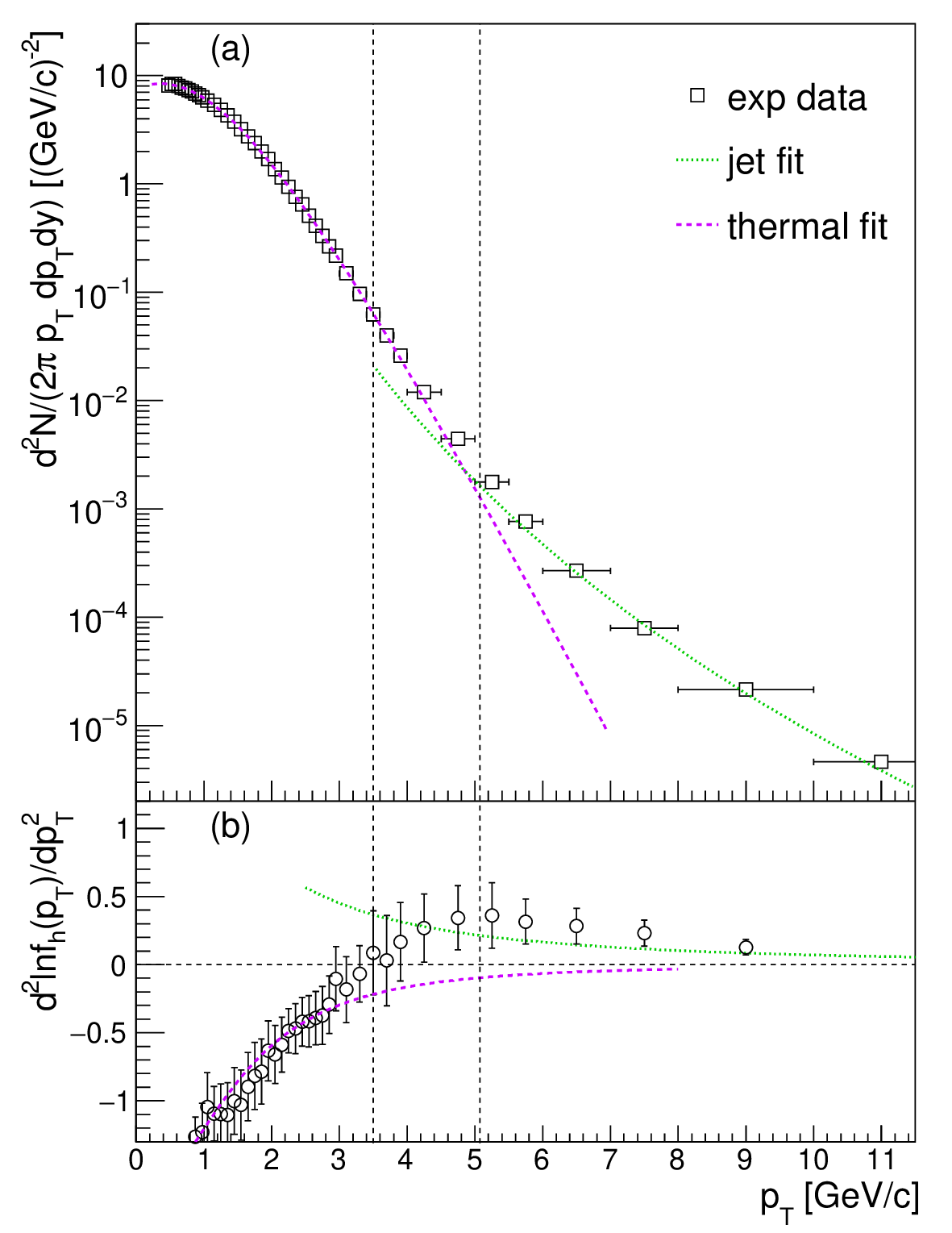}
\par\end{centering}
\caption{(a) $p_{T}$ spectrum of proton at mid-rapidity in 0-5\% centrality
in Pb-Pb collisions at $\sqrt{s_{NN}}=2.76$ TeV \citep{ALICE:2014juv}
and two fitting lines; (b) the second derivative of the logarithm
of the spectrum and two fitting functions with respect to $p_{T}$.
\label{fig:p_c05}}
\end{figure}

Fig.~\ref{fig:p_c05} shows the similar analysis for $p_{T}$ spectrum
of proton in central Pb-Pb collisions at $\sqrt{s_{NN}}=2.76$
TeV. Functions in Eqs.~(\ref{eq:fpt_thermal}) and (\ref{eq:fpt_jet})
can well fit the data in low and large $p_{T}$ range, respectively.
Levy-Tsallis function in Eq.~(\ref{eq:fpt_levy_v0}) can not globally
fit the data very well because of strong collective flow in central
collisions. Taking advantage of the precise measured data, we alternatively
carry out a local fit with an exponentially third-order polynomial function 
(see Sec. \ref{sec:ana_method} for details) and obtain the second
derivative of the logarithm of proton spectrum. Results are shown
in bottom panel as open circles. With the help of two vertical guiding
lines crossing two panels and fits of Eqs.~(\ref{eq:fpt_thermal})
and (\ref{eq:fpt_jet}), we see that the spectrum of proton in central
collisions can also be clearly split into three regions. In central
collisions, $p_{0}$ point is about 3.5 GeV/$c$ and $p_{1}$ point is
about 5.0 GeV/$c$, and both are larger than those in peripheral collisions
about 1.5 GeV/$c$. This movement is because of the stronger collective
flow in central collisions. We also note that the width of transition
region $p_{1}-p_{0}$ is about 1.5 GeV/$c$ in both central and peripheral
collisions. From above two examples, we see that $p_{0}$ can serve
as the characteristic point of locating the domain (up limit) of soft/thermal
protons in transverse momentum space and $p_{1}$ can serve as 
the characteristic point of locating the domain (low limit) of hard/jet
protons in transverse momentum space.

For other hadrons such as $K$, $\phi$ and strange baryons in heavy-ion
collisions and these hadrons in $pp$ and $p$-Pb collisions at above
and other collision energies, we can also see the clear division of
three kinetic regions like Figs.~\ref{fig:p_c6080} and \ref{fig:p_c05}
and find the movement of $p_{0}$ and $p_{1}$ points for different
hadron species, collision systems and collision energies. Therefore,
in this paper, we systematically study the dependence of $p_{0}$
and $p_{1}$ points on hadron species, collision system and collision
energy and discuss the properties of two characteristic points.
Furthermore, we focus on the property of soft hadrons by studying
the correlation between $p_{0}$ and the average transverse momentum
($\left\langle p_{T}\right\rangle $) of hadrons in $pp$, $p$-Pb
and heavy-ion collisions at RHIC and LHC energies.  

The paper is organized as follows. Sec. \ref{sec:ana_method} introduces
the data analysis method and solution of $p_{0}$ and $p_{1}$ points.
Sec. \ref{sec:p0_p1_results} shows the results of obtained $p_{0}$
and $p_{1}$ points of identified hadrons in $pp$, $p$-Pb and heavy-ion
collisions at RHIC and LHC energies. Sec. \ref{sec:p0_apt_corr} discusses
the correlation between $p_{0}$ of hadrons and their $\left\langle p_{T}\right\rangle $.
Sec. \ref{sec:summary} gives the summary. 
\hl{Appendix \ref{fit_vars_collect} presents results of parameter values of Eqs.~(\ref{eq:fpt_levy_v0}-\ref{eq:fpt_jet}) and (\ref{eq:fpt_levy_v2}-\ref{eq:fpt_levy_v3}) in fitting experimental data of hadronic $p_T$ spectra in high energy collisions, which are necessary for the calculation of $p_0$ and $p_1$.}

\section{analysis method of hadronic $p_{T}$ spectra\label{sec:ana_method}}

In this section, we introduce the fit method of $p_{T}$ spectra of
hadrons and calculation of characteristic points $p_{0}$ and $p_{1}$.
In order to fit $p_{T}$ spectra of hadrons with least bias, we consider
different fit functions. Besides the popular Levy-Tsallis function
in Eq.~(\ref{eq:fpt_levy_v0}), we consider its two variants
\begin{align}
\frac{dN^{Levy-v2}}{2\pi p_{T}dp_{T}} & =Np_{T}\left[1+\frac{\left(m_{T}-m\right)^{a}}{nc}\right]^{-n},\label{eq:fpt_levy_v2}\\
\frac{dN^{Levy-v3}}{2\pi p_{T}dp_{T}} & =Np_{T}\left[1+\frac{\left(m_{T}-p_{T}v\right)/\sqrt{1-v^{2}}}{nc}\right]^{-n}
    \label{eq:fpt_levy_v3}
\end{align}
by adding parameter $a$ in $Levy-v2$ and $v$ in $Levy-v3$ to more
freely tune the shape of $p_{T}$ spectrum. 

In globally fitting data of hadronic $p_{T}$ spectra using Eqs.~(\ref{eq:fpt_levy_v0})
and (\ref{eq:fpt_levy_v2}-\ref{eq:fpt_levy_v3}), we firstly transform
the data of hadronic spectrum to the form of $dN/dp_{T}$ and then
perform the fit procedure. $\chi^{2}$ is evaluated by
\begin{equation}
\chi^{2}=\frac{1}{N_{data}}\sum_{i=1}^{N_{data}}\left(\frac{F(p_{T,i})-y_{i}}{\sigma_{i}}\right)^{2}.
\end{equation}
Here, $y_{i}$ is the center value of experimental datum for the yield
density $dN/dp_{T}$ at point $p_{T,i}$ and $\sigma_{i}$ is the
statistical error of yield density. $F(p_{T,i})=\left(\int_{p_{T,i}-\Delta p_{T}/2}^{p_{T,i}+\Delta p_{T}/2}2\pi p_{T}f(p_{T})dp_{T}\right)/\Delta p_{T}$
is the yield density calculated by the fit function via the integral
over bin $i$ with width $\Delta p_{T}$. Parameters of fit function
are obtained by the minimum of $\chi^{2}$ with non-linear least square
method. Sometimes, data of hadronic $p_{T}$ spectra contain the point
of relatively high $p_{T}$. Because the purpose of fitting hadronic
$p_{T}$ spectra is to find the characteristic points $p_{0}$ and
$p_{1}$ which are not very large in usual evaluation, we will neglect
experimental datum points of $p_{T}$ much higher than $p_{1}$ in
order to obtain the better quality in above global fitting. In general,
we set the maximum of $p_{T}$ of used datum point is less than 8
GeV/$c$ for baryons and 6 GeV/$c$ for mesons. 

\begin{figure}[H]
\begin{centering}
\includegraphics[scale=0.35]{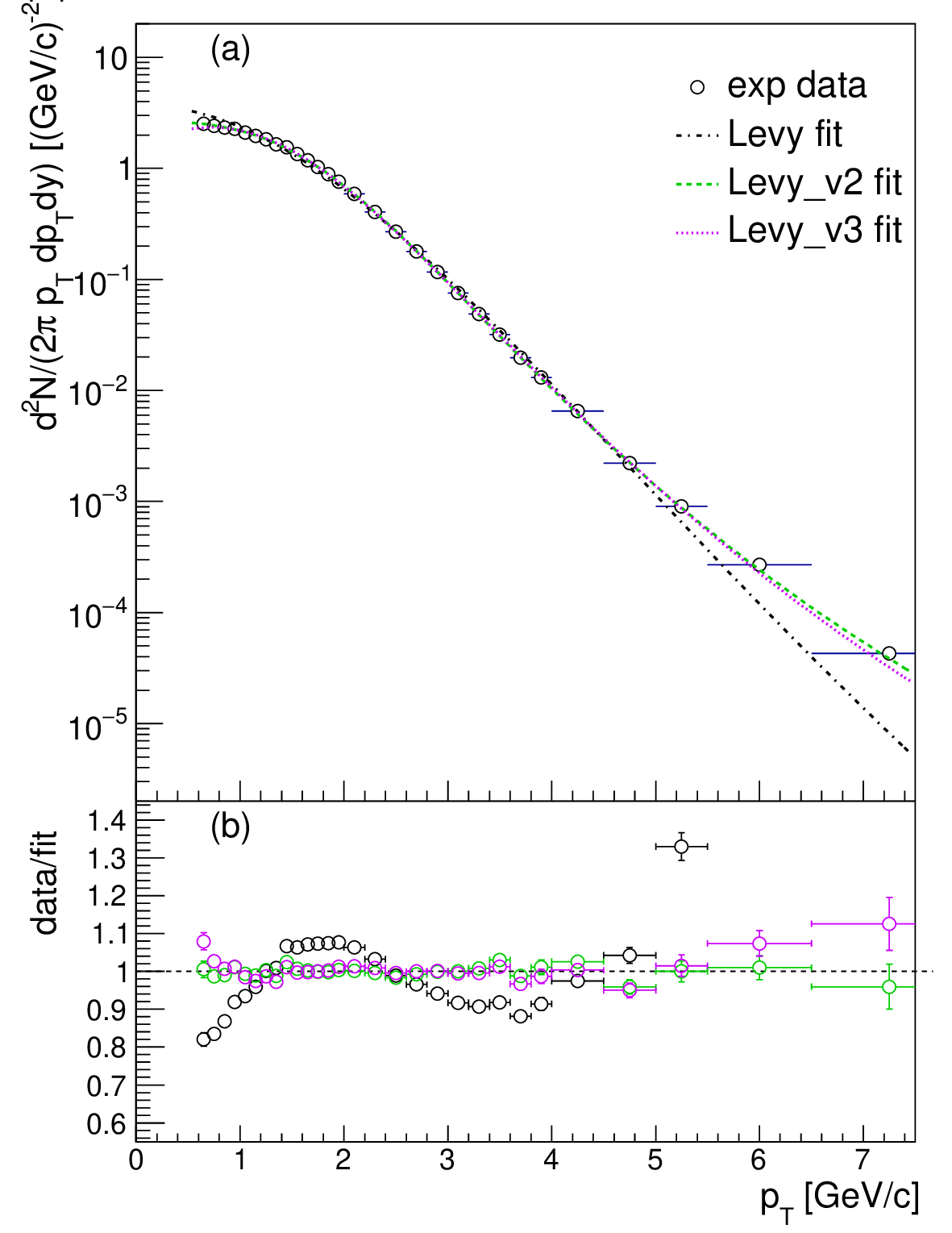}
\par\end{centering}
\caption{(a) $p_{T}$ spectrum of $\Lambda$ at mid-rapidity in 0-5\% centrality in Pb-Pb collisions at $\sqrt{s_{NN}}=2.76$ TeV \citep{ALICE:2013cdo} and three global fittings with Eq.~(\ref{eq:fpt_levy_v0}) labeled \textquotedblleft Levy fit\textquotedblright, Eq.~(\ref{eq:fpt_levy_v2}) labeled \textquotedblleft Levy\_v2 fit\textquotedblright{} and Eq.~(\ref{eq:fpt_levy_v3}) labeled \textquotedblleft Levy\_v3 fit\textquotedblright ; (b) Ratios of data to fittings. \label{fig:Lam_3fit}}
\end{figure}

In $pp$, $p$-Pb and peripheral Pb-Pb and Au+Au collisions, Levy-Tsallis
function in Eq.~(\ref{eq:fpt_levy_v0}) and its two variants in Eqs.~(\ref{eq:fpt_levy_v2}) and (\ref{eq:fpt_levy_v3}) usually well fit
the experimental data of hadronic $p_{T}$ spectrum, and the latter
two are slightly better than the standard Levy-Tsallis function. In
semi-central and central Pb-Pb collisions, Levy-Tsallis function fit
is sometimes not so good but its two variants in Eqs.~(\ref{eq:fpt_levy_v2})
and (\ref{eq:fpt_levy_v3}) are usually fine. Fig.~\ref{fig:Lam_3fit}
is an example of this case in fitting $p_{T}$ spectrum of $\Lambda$
at mid-rapidity in 0-5\% centrality in Pb-Pb collisions at $\sqrt{s_{NN}}=2.76$
TeV. 

\hl{By fitting data of hadronic $p_T$ spectra with functions in Eqs.(\ref{eq:fpt_levy_v0}) and (\ref{eq:fpt_levy_v2}-\ref{eq:fpt_levy_v3}), we obtain the corresponding parameters in three functions. Then we evaluate $\left[\ln f_{h}\left(p_{T}\right)\right]^{''}$ according to the definition in Eq.~(\ref{eq:lnfpt_p2_def}) and apply a Newton method to numerically solve $\left[\ln f_{h}\left(p_{T}\right)\right]^{''}=0$ to obtain the root $p_{0}$ and also apply the Newton method to find the maximum of $\left[\ln f_{h}\left(p_{T}\right)\right]^{''}$ and record its $p_T$ position as $p_{1}$.
}
The errors of $p_{0}$ and $p_{1}$ are calculated
by the propagation of errors of parameters in the fit function. Final
results of $p_{0}$ and $p_{1}$ are the weighted average of above
three fittings
\begin{equation}
p_{0,1}=\sum_{i}\omega_{i}p_{0,1}^{(i)}\label{eq:p01_eval}
\end{equation}
where index $i$ runs over three sets $Levy$, $Levy\_v2$ and $Levy\_v3$
and weight factor $\omega_{i}$ is $\chi^{-2}$ in $i$-set fit normalized
by the summation of $\chi^{-2}$ of three sets. In special case such
as that shown in Fig.~\ref{fig:Lam_3fit}, we can drop the results
of Levy-Tsallis fit and take the average values of latter two fit
functions or consider the results of re-fitting in the constrained
$p_{T}$ range as discussed below. 

Besides the above global fitting method, we also try the local fit
of experimental data of hadronic $p_{T}$ spectra. Since we expect
that $\left[\ln f_{h}\left(p_{T}\right)\right]^{''}$ should be a
continuous and smoothly-changed function, we use the exponential of
three-order polynomial $\exp\left[P_{3}(p_{T})\right]$ as the local
fit function. This fit function can capture the change of $\left[\ln f_{h}\left(p_{T}\right)\right]^{''}$
at the given $p_{T}$ point and meanwhile avoid the large fluctuation
of $\left[\ln f_{h}\left(p_{T}\right)\right]^{''}$ in the vicinity
of $p_{T}$ (which will occur if higher-order polynomial is used).
The fitting procedure is as follows. First, we select a $p_{T,i}$
according to the bin center value of a datum point indexed $i$. Second,
we select the $p_{T}$ range for $\exp\left[P_{3}(p_{T})\right]$
fit which includes $p_{T,i}$ point. Because $P_{3}$ polynomial has
4 parameters, the selected range includes at least 4 points and we
usually select the window $[p_{T,i-1}-\Delta p_{T,i-1}/2,p_{T,i+2}+\Delta p_{T,i+2}/2]$
if data in this range are available. Datum points in the low $p_{T}$
range are very dense where bin width $\Delta p_{T,i}$ is very small
(e.g. 0.1 for Fig.~\ref{fig:p_c05}). In this case the fluctuation
of any point in $[p_{T,i-1},p_{T,i+2}]$ will cause large fluctuation
of $\left[\ln f_{h}\left(p_{T}\right)\right]^{''}$ which is disadvantageous
to obtain the smoothly-changed derivatives. Therefore, in this case
we can properly extend the range $[p_{T,i}-w,p_{T,i}+w]$ with $w>0.5$
GeV/$c$ to obtain the relatively smoothly-changed $\left[\ln f_{h}\left(p_{T}\right)\right]^{''}$.
The result in panel (b) in Fig.~\ref{fig:p_c05} is an example of
local fit with $\exp\left[P_{3}(p_{T})\right]$ function. We see that
the obtained $\left[\ln f_{h}\left(p_{T}\right)\right]^{''}$for $p_{T}\lesssim4$
GeV/$c$ has a certain fluctuation, which is because of the very dense
datum points in the low $p_{T}$ range. As $p_{T}\gtrsim4$ GeV/$c$,
datum points are relatively sparse and therefore relatively large $p_{T}$
interval among datum points will cause the relatively small error
of $\left[\ln f_{h}\left(p_{T}\right)\right]^{''}$.

The above $\exp\left[P_{3}(p_{T})\right]$ local fit method is only
auxiliary method in our practical calculations. Actually, Eqs.~(\ref{eq:fpt_levy_v0})
and (\ref{eq:fpt_levy_v2}-\ref{eq:fpt_levy_v3}) can also be used
to carry out the local fit. For example, although the global fit of $\Lambda$
spectra in Fig.~\ref{fig:Lam_3fit} with Levy-Tsallis function Eq.~(\ref{eq:fpt_levy_v0}) is not very good, the re-fit of $\Lambda$ spectra in the constrained range $2.0\leq p_{T}\leq7.5$ GeV/$c$ using
Eq.~(\ref{eq:fpt_levy_v0}) can obtain the good accuracy almost the
same as those with functions Eqs.~(\ref{eq:fpt_levy_v2}-\ref{eq:fpt_levy_v3}) in above selected $p_{T}$ range. Therefore, in practical calculation
of $p_{0}$ and $p_{1}$ points by Eq.~(\ref{eq:p01_eval}), if $\chi^{2}$
of three global fittings have large difference, we will change the
$p_{T}$ range for the fit function (usually the Levy-Tsallis function)
with abnormally high $\chi^{2}$ and apply the local fit to improve
the fit quality in the vicinity of $p_{0}$ ($p_{1})$ point and finally
give the formal estimation of $p_{0}$ ($p_{1})$ point using the
improved $\chi^{2}$ fit. 

\section{results of $p_{0}$ and $p_{1}$ of identified hadrons in high energy
collisions \label{sec:p0_p1_results}}

\begin{figure}[H]
\begin{centering}
\includegraphics[width=0.85\columnwidth]{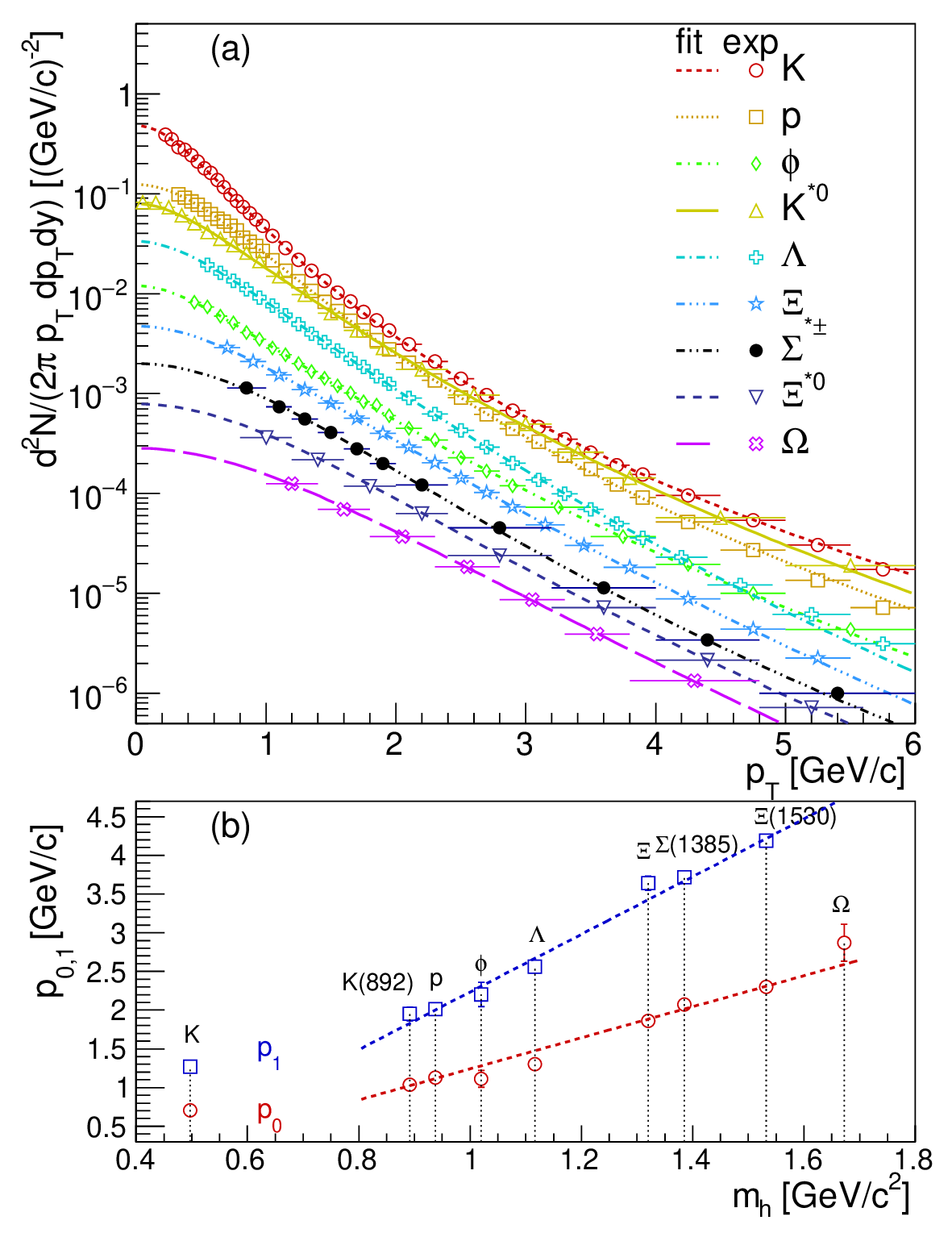}
    \caption{(a) Results (lines) of Levy-Tsallis fittings for experimental data of $p_{T}$ spectra of identified hadrons at mid-rapidity in inelastic $pp$ collisions at $\sqrt{s}=13$ TeV \citep{ALICE:2020jsh,ALICE:2023egx} and (b) the obtained $p_{0}$ (open circles) and $p_{1}$ (open squares).
\label{fig:pp13TeV_p0p1}}
\par\end{centering}
\end{figure}

In Fig.~\ref{fig:pp13TeV_p0p1}(a), we show the fittings of experimental
data for $p_{T}$ spectra of identified hadrons at mid-rapidity in
inelastic (INEL) $pp$ collisions at $\sqrt{s}=13$ TeV \citep{ALICE:2020jsh,ALICE:2023egx}.
Data of pion are not included because of too much contamination from
decays of other hadrons. Because the fit quality of three functions
Eqs.~(\ref{eq:fpt_levy_v0}) and (\ref{eq:fpt_levy_v2}-\ref{eq:fpt_levy_v3})
is all quite good, we only show results of standard Levy-Tsallis
function in Eq.~(\ref{eq:fpt_levy_v0}). In Fig.~\ref{fig:pp13TeV_p0p1}
(b), we show the obtained results of $p_{0}$ and $p_{1}$ points
for different hadrons. Generally, both $p_{0}$ and $p_{1}$ increase
with the increase of hadronic mass. Except the result of kaon, $p_{0}$
($p_{1}$) of $K^{*}$, $\phi$, $p$, $\Lambda$, $\Sigma^{*}$, $\Xi$,
$\Xi^{*}$ and $\Omega$ roughly follow the linear increase with the
mass. The width of interplay window, i.e., $p_{1}-p_{0}$, is also
increased with the hadronic mass. Here, we note that $p_{0}$ and
$p_{1}$ of proton are close to those of $K^{*}$ and $\phi$ and results
of three hadrons generally follow the mass order. This is different
from the property observed in average transverse momenta ($\left\langle p_{T}\right\rangle $)
of hadrons in high energy $pp$ collisions \citep{ALICE:2020jsh}
where $\left\langle p_{T}\right\rangle $ of proton is obviously smaller
than those of two other hadrons. 

\begin{figure}[H]
\centering{}\includegraphics[width=0.85\columnwidth]{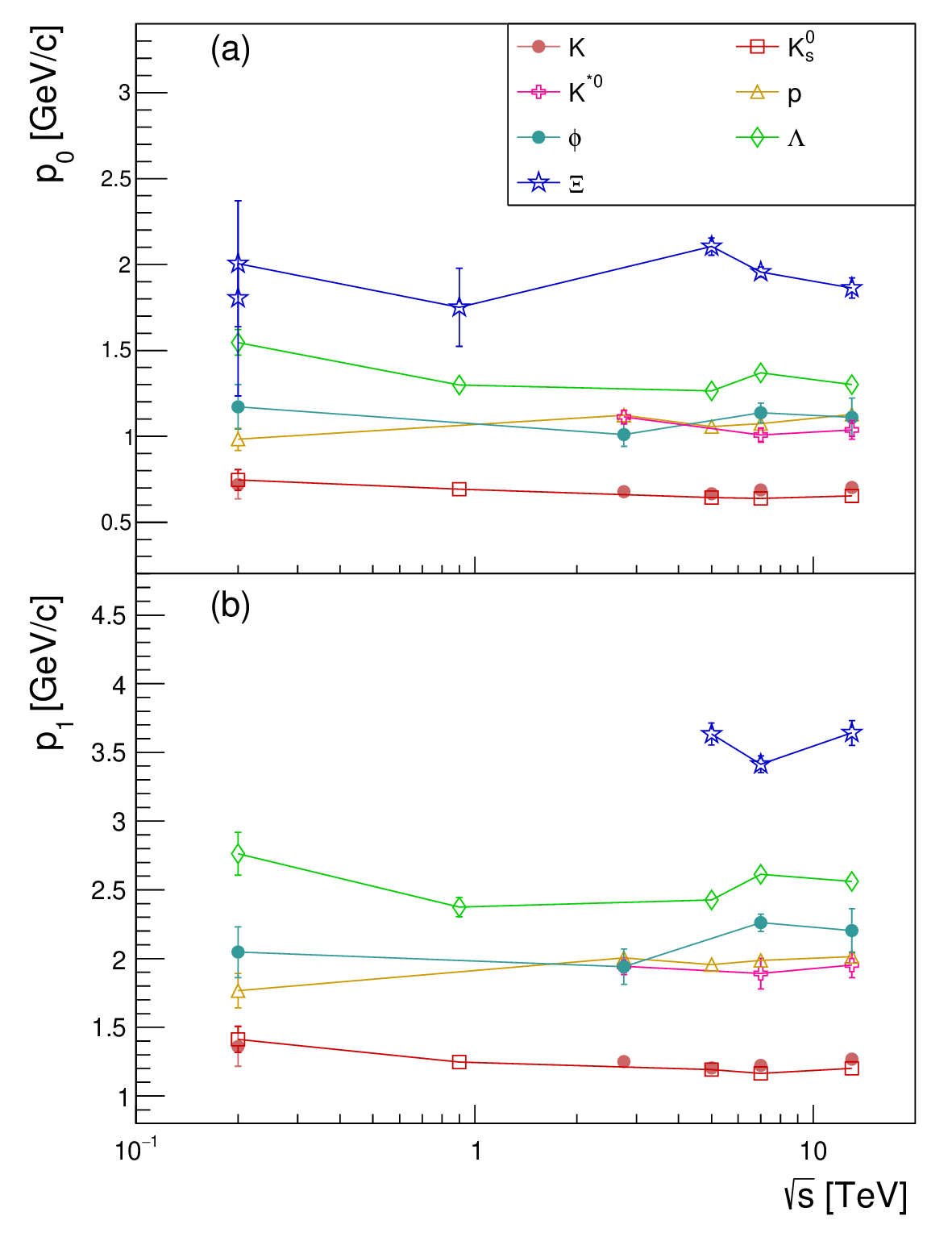}
    \caption{$p_{0}$ and $p_{1}$ of identified hadrons at mid-rapidity in inelastic $pp$ collisions at $\sqrt{s}=13,7,5.02,2.76$ TeV and those in NSD events in $pp$ collisions at $\sqrt{s}=200$ GeV and those at 900 GeV ($|y|<2$). \label{fig:p0p1_snn_dep_pp}}
\end{figure}

In Fig.~\ref{fig:p0p1_snn_dep_pp}, we show $p_{0}$ and $p_{1}$
of identified hadrons at mid-rapidity in inelastic $pp$ collisions
at $\sqrt{s}=13,7,5.02,2.76$ TeV and those in non-single diffractive (NSD) events in $pp$ collisions at $\sqrt{s}=200$ GeV and those
at 900 GeV ($|y|<2$). Experimental data for $p_{T}$ spectra of hadrons
used to fit are taken from measurements at LHC \citep{ALICE:2020jsh,ALICE:2023egx,ALICE:2021ptz,ALICE:2019hno,ALICE:2017ban,ALICE:2014juv}
and RHIC \citep{STAR:2006nmo,STAR:2006xud,CMS:2011jlm}. We see that
$p_{0}$ and $p_{1}$ of these hadrons do not exhibit a significant
dependence on collision energy. Results in $pp$ collisions at $\sqrt{s}=$200
GeV are seemingly little higher than those at LHC energies. This weak
dependence on collision energy may be related to the fact that the charged-particle
multiplicity (which can characterize the size of the soft parton system
created in collisions) in INEL(NSD) events at these collision energies
is small and is not changed dramatically in the studied energy range. 

\begin{figure}[H]
\centering{}\includegraphics[width=0.85\columnwidth]{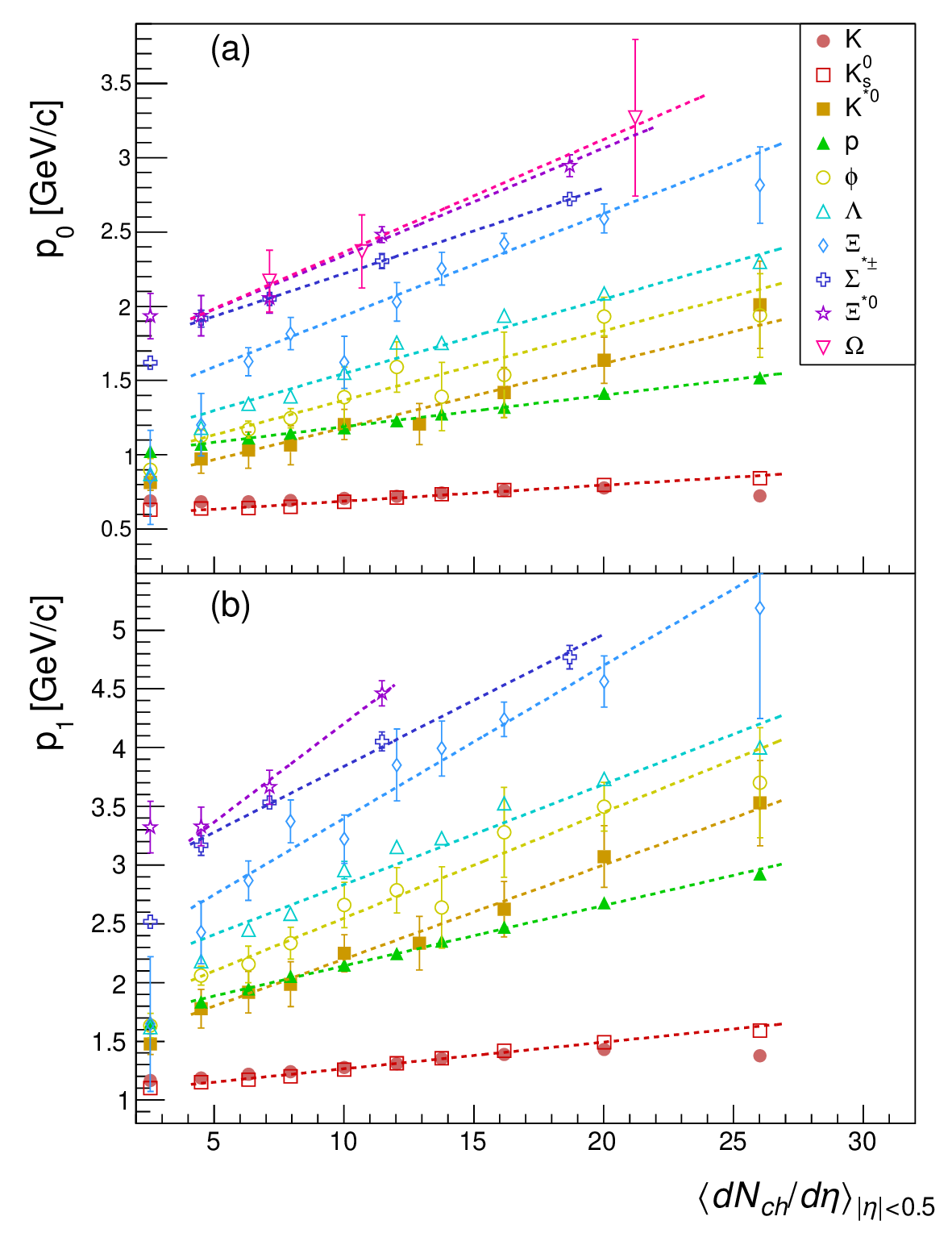}
    \caption{$p_{0}$ and $p_{1}$ points of identified hadrons at mid-rapidity in different multiplicity classes in $pp$ collisions at $\sqrt{s}=13$ TeV. \label{fig:pp13TeV_p0p1_nch_dep}}
\end{figure}

In order to study the relationship between $p_{0}$($p_{1})$ and
the size of soft parton system created in small collision systems,
we show results of $p_{0}$ and $p_{1}$ for identified hadrons in different multiplicity event classes in $pp$ collisions at $\sqrt{s}=13$ TeV in Fig.~\ref{fig:pp13TeV_p0p1_nch_dep}. The experimental
data of hadrons used in fit are taken from \citep{ALICE:2023egx,ALICE:2020nkc,ALICE:2019etb,ALICE:2019avo}.
We see that both $p_{0}$ and $p_{1}$ obviously increase with the
increase of the multiplicity density of charged particles at mid-rapidity.
This increase suggests the expansion of soft interactions for hadron
production in momentum space in high multiplicity events. In order
to better visualize this increase, we present the linear fittings
for results of $p_{0}$ and $p_{1}$ of different hadrons and show
them as dashed lines in Fig.~\ref{fig:pp13TeV_p0p1_nch_dep}. We see
that the linear fit basically captures the increasing feature of $p_{0}$
and $p_{1}$ in the multiplicity range of $5\lesssim\left\langle dN_{ch}/d\eta\right\rangle _{|\eta|<0.5}\lesssim25$.
In addition, from these fitting lines, we see that the increase of
$p_{0}$ ($p_{1}$) with multiplicity is dependent on the hadron species.
Generally, the increase rate (i.e., the slope of the fitting line)
is proportional to the mass of hadron. There is one exception
that the increase rate of proton (see the up-triangles and the fitting
line) is smaller than that of $K^{*}$ (see solid squares and the
fitting line). 

\begin{figure}[H]
\centering{}\includegraphics[width=0.85\columnwidth]{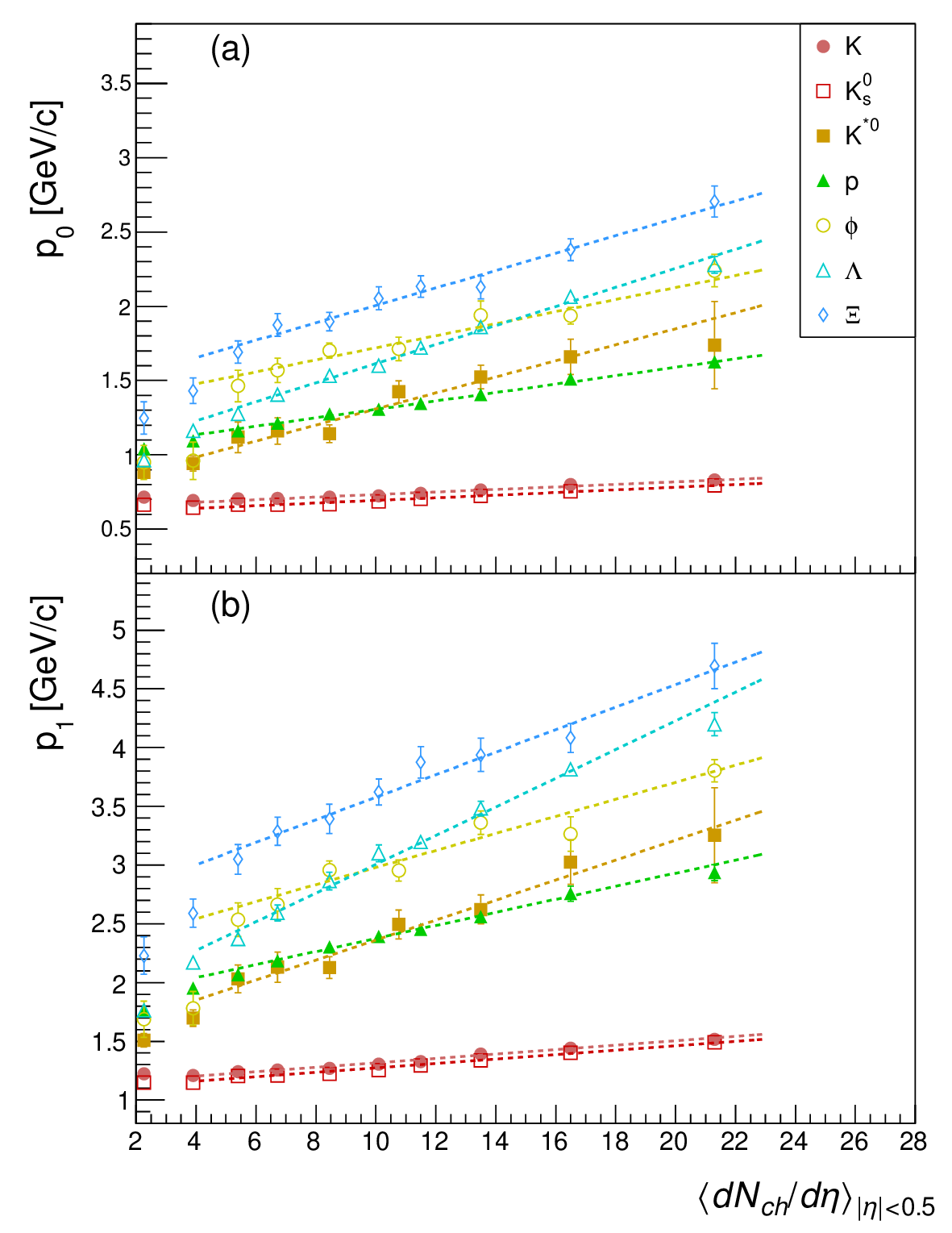}
    \caption{$p_{0}$ and $p_{1}$ points of identified hadrons at mid-rapidity in different multiplicity classes in $pp$ collisions at $\sqrt{s}=7$ TeV. \label{fig:pp7TeV_p0p1_nch_dep}}
\end{figure}

\begin{figure}[H]
\centering{}\includegraphics[width=0.85\columnwidth]{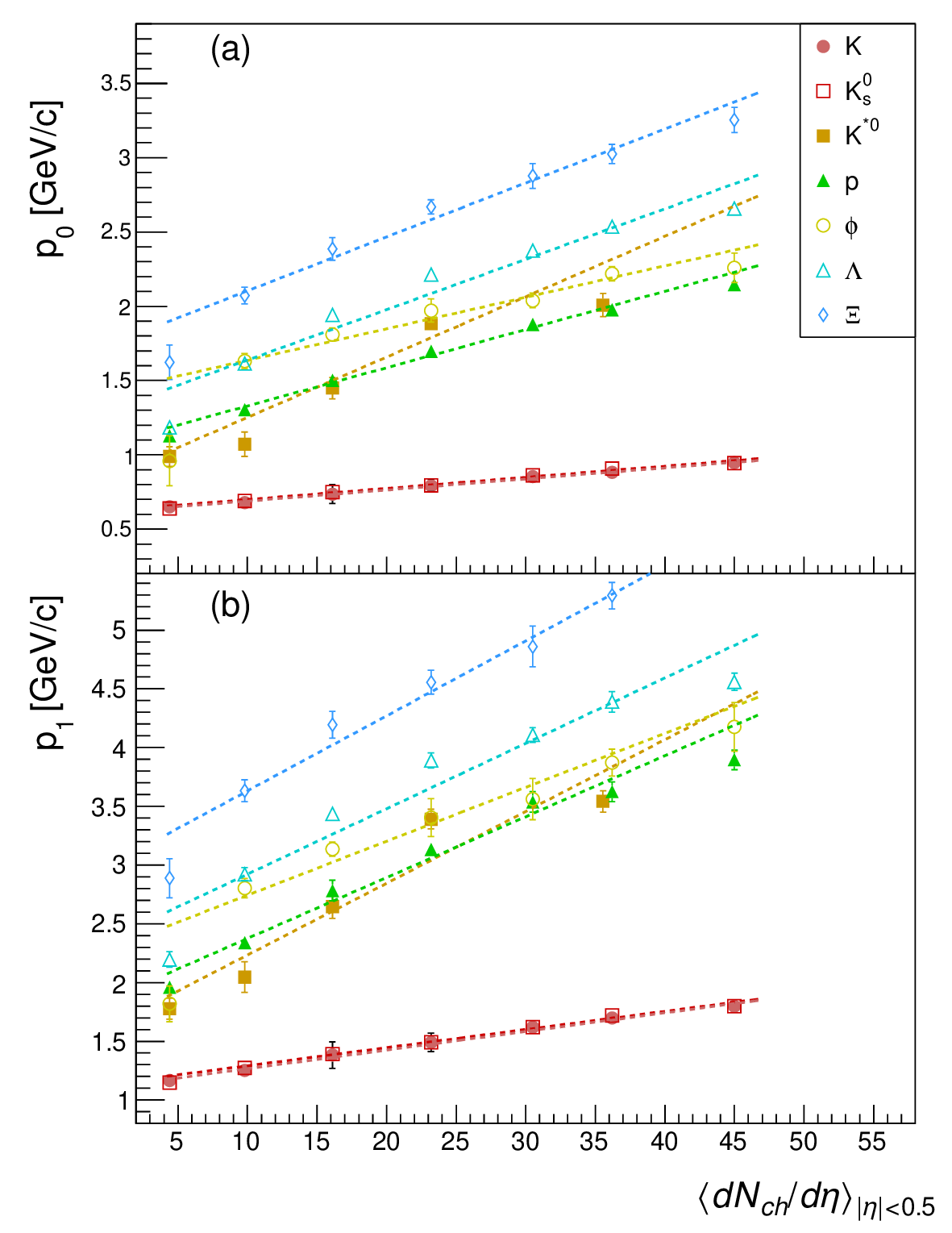}
    \caption{$p_{0}$ and $p_{1}$ points of identified hadrons at mid-rapidity in different multiplicity classes in $p$-Pb collisions at $\sqrt{s_{NN}}=5.02$ TeV. \label{fig:pPb5TeV_p0p1_nch_dep}}
\end{figure}

Using experimental data of ALICE collaboration \citep{ALICE:2020jsh,ALICE:2016fzo,ALICE:2018pal,ALICE:2016dei,ALICE:2013wgn,ALICE:2016sak},
we further calculate $p_{0}$ and $p_{1}$ of identified hadrons in
different multiplicity classes in $pp$ collisions at $\sqrt{s}=$
7 TeV and $p$-Pb collisions at $\sqrt{s_{NN}}=$ 5.02 TeV and show
results in Fig.~\ref{fig:pp7TeV_p0p1_nch_dep} and \ref{fig:pPb5TeV_p0p1_nch_dep},
respectively. $p_{0}$ and $p_{1}$ of identified hadrons are all
increased with the increase of system multiplicity, which is similar
to that in $pp$ collisions at $\sqrt{s}=13$ TeV. In particular,
we see from Fig.~\ref{fig:pPb5TeV_p0p1_nch_dep} that the linear increase
of $p_{0}$ of hadrons in $p$-Pb collisions still roughly holds in
a wider multiplicity range $5\lesssim\left\langle dN_{ch}/d\eta\right\rangle _{|\eta|<0.5}\lesssim45$.
In the highest multiplicity class, $p_{0}$ and $p_{1}$ of identified
hadrons in $p$-Pb collisions are clearly higher than those in $pp$
collisions. 

Summarizing above results, we see an obvious increase of $p_{0}$
(and $p_{1}$) of hadrons in high multiplicity events of $pp$ and
$p$-Pb collisions. This can be properly understood. The charged-particle
multiplicity is a physical quantity built from hadrons of low $p_{T}$.
Most of low $p_{T}$ hadrons are hadronization outcomes of soft parton
system produced in collisions. Higher multiplicity means soft parton
system with larger size before hadronization and means more intense
soft parton interactions in system evolution, more momentum
generation and finally stronger expansion of soft hadrons in momentum
space. Following this idea, we can expect stronger expansion of soft
hadrons in relativistic heavy-ion collisions where the size of soft
quark system is huge and the strong collective flow is formed. 

\begin{figure*}[t]
\centering{}\includegraphics[width=0.85\textwidth]{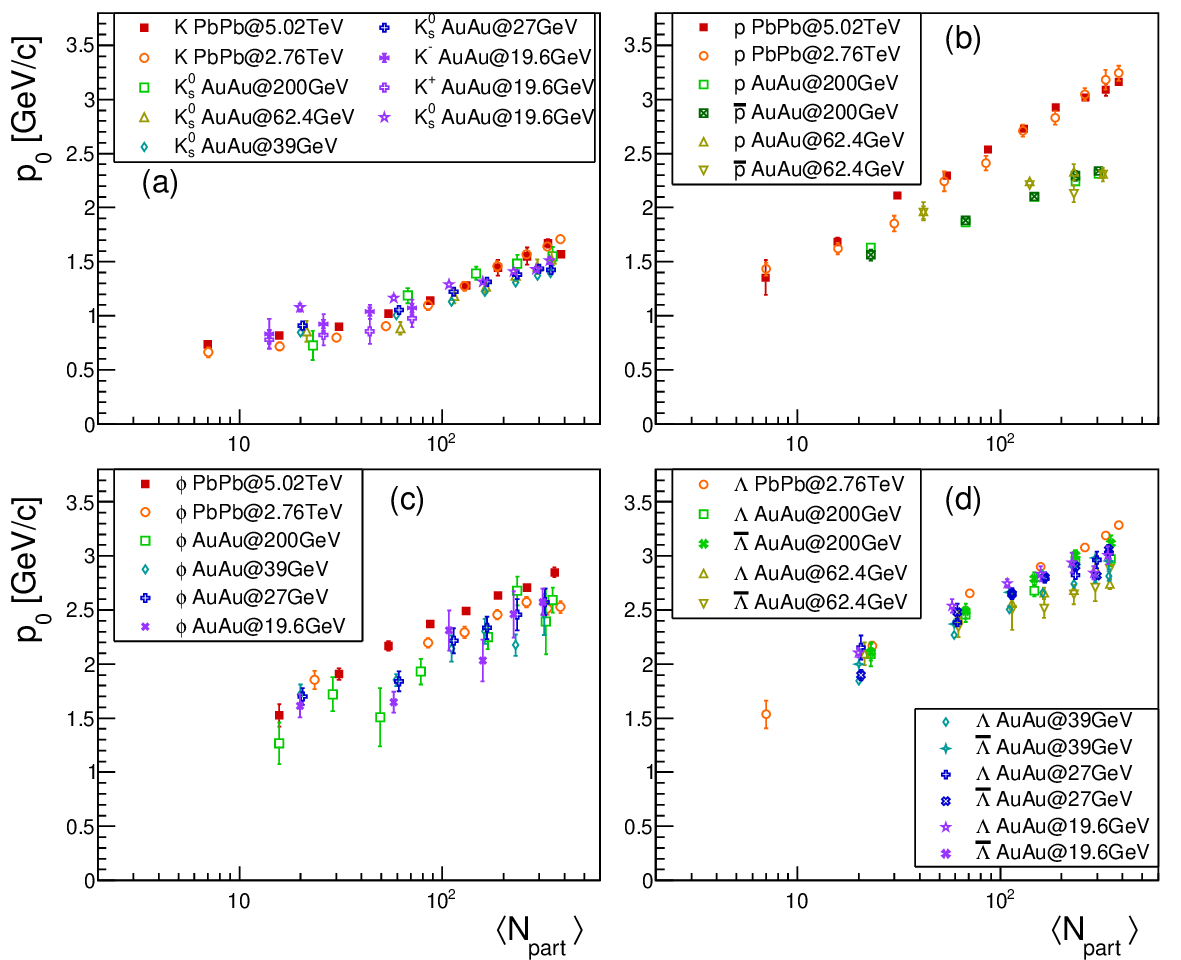}
    \caption{$p_{0}$ of hadrons at mid-rapidity in Au+Au collisions at RHIC energies and Pb-Pb collisions at LHC energies. \label{fig:p0_AA_npart}}
\end{figure*}

In Fig.~\ref{fig:p0_AA_npart}, we show systematic calculations
for $p_{0}$ of kaon, proton, $\phi$ and $\Lambda$ in different
centralities in Au+Au collisions at $\sqrt{s_{NN}}=19.6,27,39,62.4,200$
GeV and Pb-Pb collisions at $\sqrt{s_{NN}}=2.76$ and 5.02 TeV. Experimental
data of these hadrons are taken from \citep{STAR:2017sal,STAR:2019bjj,ALICE:2013mez,ALICE:2017ban,ALICE:2013cdo,ALICE:2019hno,ALICE:2019xyr}.
Because datum points at high $p_{T}$ for (strange) baryons are not rich, the calculation of $p_{1}$  is usually not complete and therefore not shown in this paper. There are two main collision parameters that influence results of $p_{0}$, i.e., collision energy and collision centrality characterized by $N_{part}$. In general, we see that $p_{0}$ of hadrons increase with the increase of $\left\langle N_{part}\right\rangle $ at each collision energy. $p_{0}$ of hadrons at high $\left\langle N_{part}\right\rangle $ are significantly larger than those in $pp$ and $p$-Pb collisions shown in Figs.~\ref{fig:pp13TeV_p0p1_nch_dep}-\ref{fig:pPb5TeV_p0p1_nch_dep}, which is mainly because of the strong collective flow in central collisions that can boost soft hadrons to larger $p_{T}$. From results of kaon in Fig.~\ref{fig:p0_AA_npart}(a) which have more datum points at low $\left\langle N_{part}\right\rangle $, we see that $p_{0}$ of kaon in peripheral collisions are close to those in $pp$ and $p$-Pb collisions at LHC energies as shown in Figs.~\ref{fig:pp13TeV_p0p1_nch_dep}-\ref{fig:pPb5TeV_p0p1_nch_dep}.  Result points of proton, $\phi$ and $\Lambda$ at small $\left\langle N_{part}\right\rangle $
are relatively rare and the available results are also shown to be
close to those in low multiplicity event classes in $pp$ and $p$-Pb
collisions shown in Figs.~\ref{fig:pp13TeV_p0p1_nch_dep}-\ref{fig:pPb5TeV_p0p1_nch_dep}.
Fig.~\ref{fig:p0_AA_npart}(a) also shows that the increase speed
of $p_{0}$ of kaon in heavy-ion collisions at low $\left\langle N_{part}\right\rangle \lesssim40$
is relatively smaller than that at higher $\left\langle N_{part}\right\rangle $.
This may be due to the rapid increase of the collective flow for the
hot medium produced in (semi-)central collisions. In addition, we
find that $p_{0}$ of hadrons at similar $\left\langle N_{part}\right\rangle $
is generally higher at higher collision energy but this collision
energy dependence is relatively weak in comparing with their $\left\langle N_{part}\right\rangle $
dependence. 

\section{Correlation between $p_{0}$ and $\left\langle p_{T}\right\rangle $\label{sec:p0_apt_corr}}

From above results, we see that the point $p_{0}$, the zero point of the second derivative of the logarithm of invariant transverse spectrum of hadrons, shows the explicit dependence on the hadron species, collision energy and collision system. Therefore, this quantity can be used to characterize the influence of soft parton interaction in hadron production in momentum space. In this section, we compare the results of $p_{0}$ of identified hadrons with their average transverse momenta $\left\langle p_{T}\right\rangle $ and study their correlation property. Because $\left\langle p_{T}\right\rangle $ is also dominated by hadrons of low $p_{T}$, the combination study of $\left\langle p_{T}\right\rangle $ and $p_{0}$ can give us more information of soft hadron production in momentum space. 

\begin{figure}[H]
\centering{}\includegraphics[width=0.85\columnwidth]{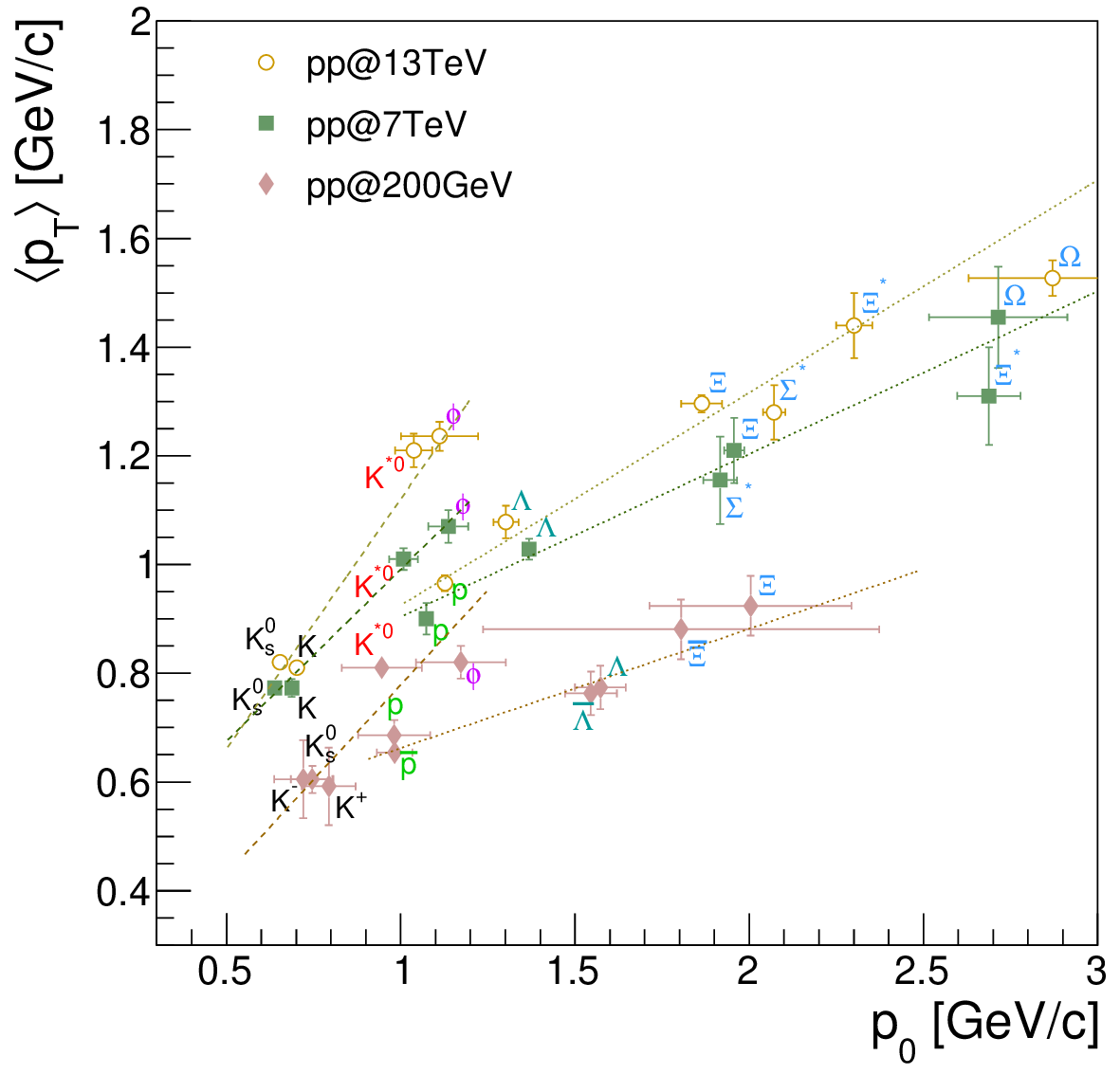}
    \caption{Correlation between $p_{0}$ and $\left\langle p_{T}\right\rangle $ for hadrons at mid-rapidity in INEL events in $pp$ collisions at $\sqrt{s}=13,7$ TeV and NSD events in $pp$ collisions at $\sqrt{s}=$
200 GeV. \label{fig:p0_apt_corr_pp_inel}}
\end{figure}

First, we show the correlation between $p_{0}$ and $\left\langle p_{T}\right\rangle$
of identified hadrons at mid-rapidity in INEL events in $pp$ collisions
at $\sqrt{s}=13,7$ TeV and NSD events in $pp$ collisions at $\sqrt{s}=$200
GeV in Fig.~\ref{fig:p0_apt_corr_pp_inel}. These three collision systems create parton systems of small
size. We see that $p_{0}$ of hadrons globally show positive correlation
with their $\left\langle p_{T}\right\rangle $. In comparison with
the increase of $\left\langle p_{T}\right\rangle $ about 1 GeV/$c$
from kaon to $\Omega$, the changed magnitude of $p_{0}$ exceeds 2
GeV/$c$, showing more sensitivity to hadron mass or species. In addition,
we find a special case for results of proton, $K^{*}$ and $\phi$
with close masses. We see that $p_{0}$ of proton is close to those of $K^{*}$ and $\phi$ but $\left\langle p_{T}\right\rangle $ of
proton is obviously lower than those of $K^{*}$ and $\phi$. This
indicates the production of three hadrons in small parton system created
in $pp$ collisions should be related to not only their masses but
also their baryon/meson feature or, in other words, their quark composition
structure. Inspired by this indication, we present some fitting lines
for results of mesons and results of baryons, respectively, from which
we can obtain better classification for results of these hadrons at
three collision energies. 

\begin{figure*}[t]
\centering{}\includegraphics[width=0.95\textwidth]{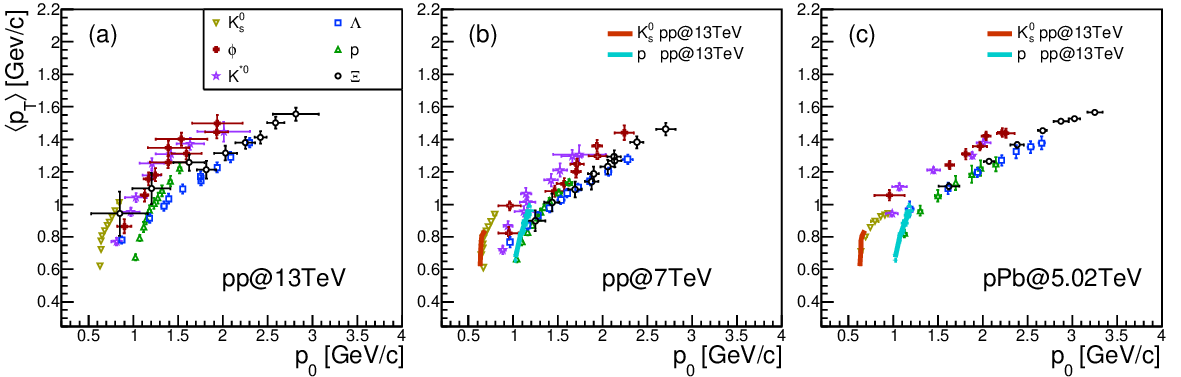}
    \caption{Correlation between $p_{0}$ and $\left\langle p_{T}\right\rangle $ of hadrons at mid-rapidity in $pp$ collisions at $\sqrt{s}=13,7$ TeV and $p$-Pb collisions at $\sqrt{s_{NN}}=5.02$ TeV. \label{fig:p0_apt_corr_nch_pp_pA}}
\end{figure*}

In Fig.~\ref{fig:p0_apt_corr_nch_pp_pA}, we show the correlation between $p_{0}$ and $\left\langle p_{T}\right\rangle $ of identified hadrons at mid-rapidity in different multiplicity event classes in $pp$ collisions at $\sqrt{s}=13,7$ TeV and in $p$-Pb collisions at $\sqrt{s_{NN}}=5.02$ TeV. We see that $p_{0}$ of hadrons are generally positively correlated with their $\left\langle p_{T}\right\rangle $ and the correlation feature is different for different hadrons. The spreading width for results of $K_{S}^{0}$ is small due to its light mass but those of $K^{*}$ and $\phi$ as well as baryons spread a relatively large region. By comparing results of $K^{*}$ and $\phi$ with those of baryons at each collision energy, we see a certain difference for $p_{0}$-$\left\langle p_{T}\right\rangle $ correlations of these hadrons. In particular, results of $p$-Pb collisions at $\sqrt{s_{NN}}=5.02$ TeV in Fig.~\ref{fig:p0_apt_corr_nch_pp_pA}(c) show a relatively clear difference between results of baryons and those of $K^{*}$ and $\phi$. This baryon/meson difference should be related to their specific production mechanism in relatively high multiplicity events in small collision systems, which may be attributed to the quark recombination mechanism of their production in high energy collisions \citep{Song:2017gcz,Gou:2017foe,Li:2021nhq}. 

We can demonstrate this baryon/meson difference by an equal-velocity
combination (EVC) model of hadron production in high energy collisions
\citep{Song:2017gcz,Gou:2017foe}. The model applies an effective
constituent quark degrees of freedom to describe the final-state parton
system just before hadronization and assumes that the hadron formation
 is dominated by the equal-velocity combination of these constituent quarks and antiquarks
\citep{Song:2017gcz}. The model can well describe the $p_{T}$ spectra
of various light-flavor hadrons and single-charm hadrons in high energy
$pp$, $p$A and AA collisions \citep{Song:2020kak,Song:2019sez,Li:2017zuj,Song:2018tpv,Song:2021ojn,Song:2023nzu}.
In one dimensional $p_{T}$ space, the $p_{T}$ spectra of baryons
and mesons $F(p_{T})\equiv dN/dp_{T}$ can be expressed as the simple
product of those of constituent quarks 
\begin{align}
F_{M}(p_{T}) & =\kappa_{M}F_{q_{1}}\left(x_{1}p_{T}\right)F_{\bar{q}_{2}}\left(x_{2}p_{T}\right),\\
F_{B}(p_{T}) & =\kappa_{B}F_{q_{1}}\left(x_{1}p_{T}\right)F_{q_{2}}\left(x_{2}p_{T}\right)F_{q_{3}}\left(x_{3}p_{T}\right)
\end{align}
where $x_{1,2}=m_{1,2}/(m_{1}+m_{2})$ for mesons and $x_{1,2,3}=m_{1,2,3}/(m_{1}+m_{2}+m_{3})$
for baryons. $m_{i}$ is the constituent mass of quark $q_{i}$. $\kappa_{M}$
and $\kappa_{B}$ are independent of $p_{T}$ and can be further decomposed
and determined \citep{Gou:2017foe,Li:2021nhq}. They are not relevant
to the derivation of $p_{0}$-$\left\langle p_{T}\right\rangle $ correlations. 

In practical calculations as we did in previous work, $p_{T}$
spectrum of quark just before hadronization can be reversely extracted
by experimental data of hadrons such as $\phi$ and proton and is
usually parameterized by the Levy-Tsallis form. Because analytical
calculation of $p_{0}$ and $\left\langle p_{T}\right\rangle $ for
Levy-Tsallis function in Eq.~(\ref{eq:fpt_levy_v0}) and its two variants
Eqs.~(\ref{eq:fpt_levy_v3}-\ref{eq:fpt_levy_v2}) is difficult,
here we take an analogous function 
\begin{equation}
f_{q}(p_{T})=A\left(1+\frac{p_{T}^{2}}{n\:c}\right)^{-n}\label{eq:levy_g2}
\end{equation}
from the extended Tsallis statistics \citep{Alemany:1994a} which
can well mimic the non-monotonic behavior of $\left[\ln f(p_{T})\right]^{''}$
such as that shown in Fig.~\ref{fig:p_c6080}. 

\begin{figure}

\begin{centering}
\includegraphics[width=0.95\columnwidth]{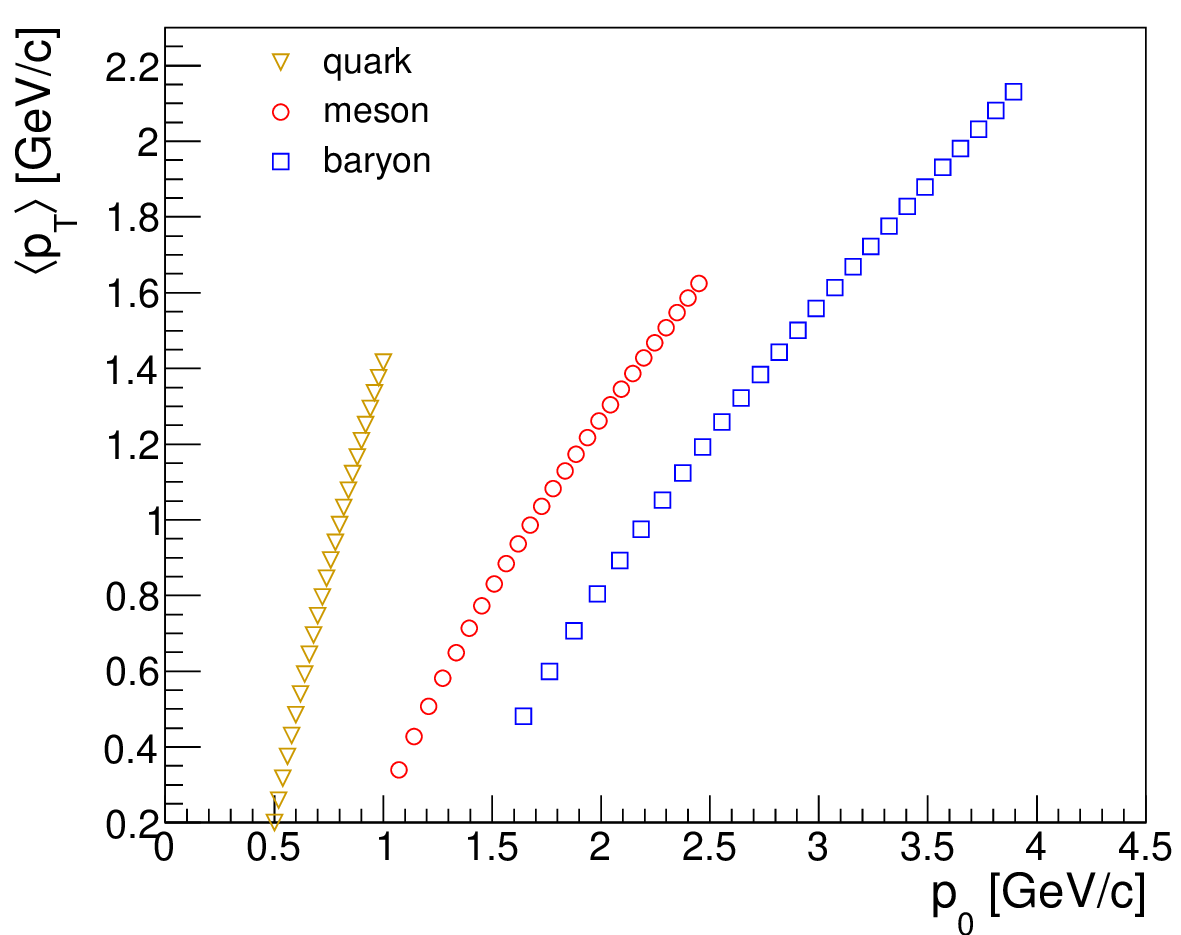}
    \caption{A schematic of the relationship among $p_{0}$-$\left\langle p_{T}\right\rangle$ correlation of quark, meson and baryon.\label{fig:po_apt_corr_evc} }
\par\end{centering}
\end{figure}

Taking the case of one quark flavor as the simplest example, we have
$F_{M}(p_{T})=\kappa_{M}F_{q}^{2}\left(p_{T}/2\right)$ and $F_{B}(p_{T})=\kappa_{B}F_{q}^{3}\left(p_{T}/3\right)$ and for invariant $p_{T}$ distribution 
\begin{align}
	f_{M}(p_{T}) & =\frac{\pi}{2}\kappa_{M} p_{T}f_{q}^{2}\left(\frac{p_{T}}{2}\right),\\
	f_{B}(p_{T}) & =\frac{4\pi^2}{27}\kappa_{B}p_{T}^{2}f_{q}^{3}\left(\frac{p_{T}}{3}\right)
\end{align}
where $F_{q}(p_{T})=F_{\bar{q}}(p_{T})$ in meson formula is assumed
for simplicity. With above three equations, we can calculate 
\begin{equation}
\left\langle p_{T}\right\rangle =\frac{\int_{0}^{\infty}p_{T}^{2}f(p_{T})dp_{T}}{\int_{0}^{\infty}p_{T}f(p_{T})dp_{T}}
\end{equation}
and solve $\left[\ln f(p_{T})\right]^{''}=0$ to get $p_{0}$ of quarks,
mesons and baryons,
\begin{align}
\left\langle p_{T}\right\rangle _{q} & =\frac{\sqrt{\pi nc}\Gamma(n-3/2)}{2\Gamma(n-1)},\\
\left\langle p_{T}\right\rangle _{M} & =\frac{4\sqrt{nc}\Gamma(2n-2)}{\sqrt{\pi}\Gamma(2n-3/2)},\nonumber \\
\left\langle p_{T}\right\rangle _{B} & =\frac{9\sqrt{\pi nc}\Gamma(3n-5/2)}{4\Gamma(3n-2)},\\
p_{0,q} & =\sqrt{nc},\\
p_{0,M} & =2\sqrt{\frac{nc\left(1+2n+2\sqrt{n(n+2)}\right)}{4n-1}},\\
p_{0,B} & =3\sqrt{\frac{nc\left(2+3n+\sqrt{3n(3n+8)}\right)}{6n-2}},
\end{align}
where $\Gamma(n)$ is Gamma function. In Fig.~\ref{fig:po_apt_corr_evc},
we show a schematic of the relationship among $p_{0}$-$\left\langle p_{T}\right\rangle $
correlation of quark, meson and baryon. We see that quark combination
produces a clear difference between $p_{0}$-$\left\langle p_{T}\right\rangle $
correlation of meson and that of baryon, which is qualitatively consistent
with the observation in Fig.~\ref{fig:p0_apt_corr_nch_pp_pA}.

In addition, when comparing $p_{0}$ of $K_{S}^{0}$ in low multiplicity event
classes at three collision energies, see the shadow areas in Fig.~\ref{fig:p0_apt_corr_nch_pp_pA} (b) and (c), we find the almost coincident
$p_{0}-\left\langle p_{T}\right\rangle $ correlations in low multiplicity
event classes at three collision energies. The same behavior also
well holds for results of proton. Results of $K^{*}$, $\phi$ and
$\Lambda$ also show this trend with relatively large statistical
uncertainties. This is an indication of the similar hadron production
for soft parton systems created in these event classes.

\begin{figure*}[t]
\centering{}\includegraphics[width=0.85\textwidth]{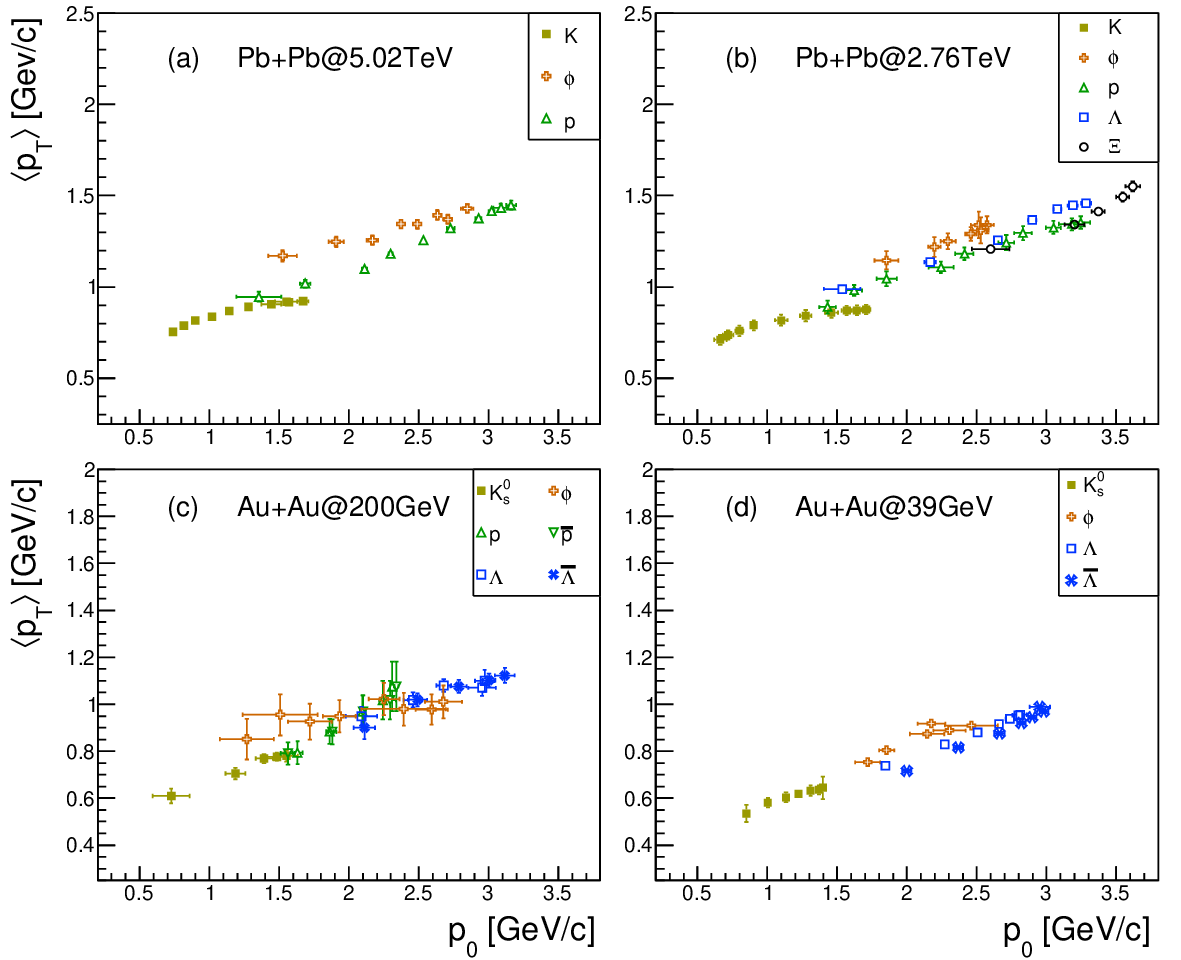}
    \caption{Correlation between $p_{0}$ and $\left\langle p_{T}\right\rangle $ for hadrons at mid-rapidity in Au+Au collisions at RHIC energies and Pb-Pb collisions at LHC energies. \label{fig:p0_apt_cor_AA}}
\end{figure*}

In Fig.~\ref{fig:p0_apt_cor_AA}, we further show the correlation
between $p_{0}$ and $\left\langle p_{T}\right\rangle $ of identified
hadrons in Pb-Pb collisions at $\sqrt{s_{NN}}=5.02,$ 2.76 TeV and
in Au+Au collisions at $\sqrt{s_{NN}}=200,$ 39 GeV. In general, we
see that $p_{0}$ of hadrons are positively correlated with their $\left\langle p_{T}\right\rangle $
and the $p_{0}$-$\left\langle p_{T}\right\rangle $ correlation feature
of hadrons shows a dependence on their species. This is similar to
that we observed in $pp$ and $p$-Pb collisions. From Fig.~\ref{fig:p0_apt_cor_AA}(b)
and (c), we see that results of proton, $\Lambda$ and $\Xi$ generally
follow a common trend, which is different from that of $\phi$
to a certain extent, and therefore this indicates a production difference
between baryon and meson in relativistic heavy-ion collisions. 

\begin{figure*}[t]
\centering{}\includegraphics[width=0.85\textwidth]{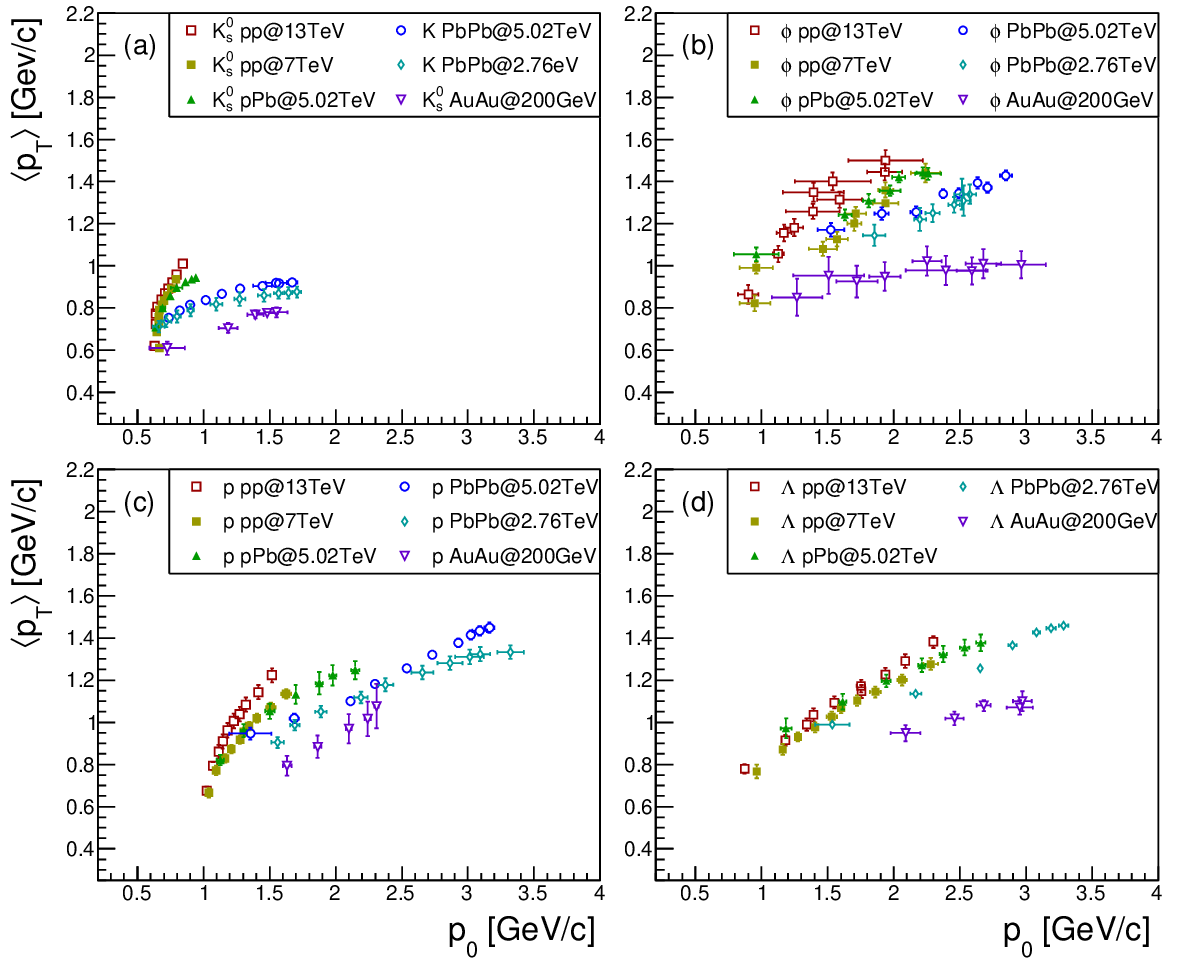}
    \caption{Comparison of $p_{0}$-$\left\langle p_{T}\right\rangle $ correlations of hadrons at mid-rapidity in high energy $pp$, $p$A and AA collisions.
\label{fig:p0_apt_cor_pp_pA_AA_com}}
\end{figure*}

In order to compare hadron production in collision systems of different
size, we show $p_{0}$-$\left\langle p_{T}\right\rangle $
correlations of identified hadrons in $pp$, $p$-Pb, Pb-Pb collisions
at LHC energies in Fig.~\ref{fig:p0_apt_cor_pp_pA_AA_com}. Results of Au+Au collisions at $\sqrt{s_{NN}}=200$
GeV are also shown as the example of low collision energy. In general,
we see that $p_{0}$ results of four hadrons in low multiplicity event
classes in $pp$, $p$-Pb collisions are close to those in peripheral
AA collisions at the studied collision energies. With the increase
of multiplicity or collision centrality, $p_{0}$-$\left\langle p_{T}\right\rangle $
correlations of identified hadrons at different collision energies
are split in the plane. Taking results of kaon ($K_{S}^{0}$) in Fig.~\ref{fig:p0_apt_cor_pp_pA_AA_com}(a) as an example, we see that the
increase of $p_{0}$ with respect to $\left\langle p_{T}\right\rangle $
in small collision systems (i.e., $pp$ and $p$-Pb collisions) is
slower than that in heavy-ion collisions. Results of $\phi$, $p$
and $\Lambda$ also show this feature. This is because the parton
systems created in heavy-ion collisions are large and evolve strong
collective flow in the expansion and cooling process, which can significantly
boost soft/thermal hadrons and corresponding $p_{0}$ to higher $p_{T}$. 

\section{Summary and discussion \label{sec:summary}}

In this paper, we analyzed the shape property of $p_{T}$ spectra of identified hadrons measured in high energy $pp$, $p$A and AA collisions and found 
\hl{ two characteristic points of the invariant $p_T$ spectra of hadrons which are helpful to empirically constrain the domain of soft hadrons and hard hadrons in transverse momentum space. }
The production of hadrons with low $p_{T}$ dominantly comes from the hadronization of soft parton system produced in collisions, and the $p_{T}$ spectrum at low $p_{T}$ tends to be exponentially decreased via Boltzmann form. 
 \hl{As illustrated by hydrodynamic model calculations \cite{Huovinen:2006jp, Kestin:2008bh,Song:2013qma, McDonald:2016vlt, Schenke:2010nt,Zhu:2015dfa}, the $\left[\ln f(p_{T})\right]^{''}$ of hadrons in low $p_T$ range is negative.} 
 The production of hadrons with high $p_{T}$ is mainly from the hadronization of high energy parton/jet, and $p_{T}$ spectra of these hadrons are usually power-law decreased. 
\hl{As illustrated by perturbative QCD calculations \citep{deFlorian:2017lwf,Albino:2008fy,deFlorian:2007ekg}, the $\left[\ln f(p_{T})\right]^{''}$ of hadrons at high $p_T$ is positive}
and decreased with $p_{T}$ . 
Finally, under a global view, we will see that $\left[\ln f(p_{T})\right]^{''}$ of hadrons at low $p_{T}$ is negative and, because of the increased contribution of hadrons from high energy jet, $\left[\ln f(p_{T})\right]^{''}$ of hadrons will rapidly increase with $p_{T}$ at intermediate $p_{T}$ and then across the zero point, maximum point and finally decrease slowly with $p_{T}$ at high $p_{T}$. 
\hl{ The zero point ($p_0$) of $\left[\ln f(p_{T})\right]^{''}$ and the maximum point ($p_1$) of $\left[\ln f(p_{T})\right]^{''}$ are two characteristic points of the invariant $p_T$ spectra of hadrons, which can classify hadrons produced in high energy collisions into three groups.  Hadrons in the region $0<p_T<p_0$ have the property $\left[\ln f(p_{T})\right]^{''}<0$ and they dominantly come from the hadronization of soft parton system created in collisions. Hadrons in the region $p_T>p_1$ have the property $\left[\ln f(p_{T})\right]^{''}>0$ and they dominantly come from the hadronization of high energy partons created in collisions.  Hadrons in the region $p_0<p_T<p_1$ are the mixing contribution of hadronization outcomes of soft parton system and high energy partons created in collisions. 
}

Using the available experimental data of hadronic $p_{T}$ spectra at RHIC and LHC, we carried out a systematic calculation for $p_{0}$ and $p_{1}$ of identified hadrons in $pp$, $p$A and AA collisions at RHIC and LHC energies, and found the following properties.
\begin{enumerate}
\item In inelastic and/or NSD events of $pp$ collisions at $\sqrt{s}=$13,
7, 5, 2.76, 0.9, 0.2 TeV, $p_{0}$ and $p_{1}$ of hadrons at the
same collision energy are dependent on their masses/species. $p_{0}$
and $p_{1}$ of $K^{*}$, $\phi$, $p$, $\Lambda$, $\Xi$, $\Sigma^{*}$,
$\Xi^{*}$, $\Omega$ roughly follow a linear increase with their
masses. $p_{0}$ and $p_{1}$ of the hadron at these collision energies
are not significantly changed. 
\item In different charged-particle multiplicity event classes in $pp$
collisions at $\sqrt{s}=$13 and 7 TeV, $p_{0}$ and $p_{1}$ of above
hadrons including kaon and $K_{S}^{0}$ show an obviously increase
with system multiplicity. The increasing behavior is roughly linear
in the multiplicity range $5\lesssim\left\langle dN_{ch}/d\eta\right\rangle _{|\eta|<0.5}\lesssim25$.
Results in $p$-Pb collisions at $\sqrt{s_{NN}}=5.02$ TeV are quite
similar and the linear increasing behavior is found in expanded multiplicity
range $5\lesssim\left\langle dN_{ch}/d\eta\right\rangle _{|\eta|<0.5}\lesssim45$. 
\item In relativistic heavy-ion collisions at RHIC and LHC energies, $p_{0}$
of hadrons show an obvious dependence on the number of participant
nucleons $\left\langle N_{part}\right\rangle $ which can characterize
the size of system created in collisions. $p_{0}$ of kaon with the most datum points shows an accelerating increase with $\left\langle N_{part}\right\rangle $
in (semi-)central collisions, which can be attributed to the formation
of strong collective flow in collisions. 
\end{enumerate}

Considering that $p_{0}$ describes the spread range of soft hadrons
and $\left\langle p_{T}\right\rangle $ quantifies the characteristic
value of soft hadrons in transverse momentum space, we further studied
the correlation between $p_{0}$ of hadrons and their $\left\langle p_{T}\right\rangle $ and found the following interesting properties.
\begin{enumerate}
\item In inelastic or NSD events of $pp$ collisions at $\sqrt{s}=$13,
7 and 0.2 TeV, $p_{0}$ of hadrons show a roughly positive correlation
with their $\left\langle p_{T}\right\rangle $, except that the result
of proton is below that of $K^{*}$ and that of $\phi$ in the $p_{0}$-$\left\langle p_{T}\right\rangle $
plane. Results of hadrons can be more clearly viewed by classifying
them into baryon group and meson group. 
\item In different charged-particle multiplicity event classes in $pp$
collisions at $\sqrt{s}=$13 and 7 TeV and $p$-Pb collisions at $\sqrt{s_{NN}}=5.02$
TeV, $p_{0}$-$\left\langle p_{T}\right\rangle $ correlations of different
hadrons follow some trend lines in the $p_{0}$-$\left\langle p_{T}\right\rangle $
plane. Correlations of $K^{*}$ and $\phi$ are generally different
from those of baryons, in particular, for results in $p$-Pb collisions,
showing a baryon/meson difference in $p_{0}$-$\left\langle p_{T}\right\rangle $ correlations of hadrons. In relativistic heavy-ion collisions, results
of $\phi$ are also different from those of proton and $\Lambda$. 
\item The difference in $p_{0}$-$\left\langle p_{T}\right\rangle $ correlation
between mesons and baryons can be qualitatively explained by an equal-velocity
combination of constituent quarks at hadronization, which is an indication
of the hadronization effect in high energy collisions. 
\end{enumerate}

Finally, $p_{0}$-$\left\langle p_{T}\right\rangle $ correlations
of kaon, $\phi$, proton, $\Lambda$ in different multiplicity events
in $pp$, $p$-Pb, Pb-Pb and Au+Au collisions are systematically compared.
Following the change of system multiplicity or collision centrality,
$p_{0}$-$\left\langle p_{T}\right\rangle $ correlations of hadrons
are split into different trend lines in the $p_{0}$-$\left\langle p_{T}\right\rangle $
plane according to hadron species and collision energy. This implies
the $p_{0}$-$\left\langle p_{T}\right\rangle $ correlations of hadrons
are sensitive to various collision parameters and therefore can be
used to reveal more detailed property of hadron production in high
energy collisions. We expect more precise data of $p_{T}$ spectra
of hadrons at LHC and RHIC are available in future for the continuous
study of hadronic $p_{0}$-$\left\langle p_{T}\right\rangle $ correlations
to gain deep insights into the property of hadron production in high
energy collisions. 
\begin{acknowledgments}
Authors thank R.Q. Wang for helpful discussions. This work is supported
by Shandong Provincial Natural Science Foundation (Grants No. ZR2025MS01). 
\end{acknowledgments}

\bibliographystyle{apsrev4-1}
\bibliography{ref}

\begin{thebibliography}{54}%
\makeatletter
\providecommand \@ifxundefined [1]{%
 \@ifx{#1\undefined}
}%
\providecommand \@ifnum [1]{%
 \ifnum #1\expandafter \@firstoftwo
 \else \expandafter \@secondoftwo
 \fi
}%
\providecommand \@ifx [1]{%
 \ifx #1\expandafter \@firstoftwo
 \else \expandafter \@secondoftwo
 \fi
}%
\providecommand \natexlab [1]{#1}%
\providecommand \enquote  [1]{``#1''}%
\providecommand \bibnamefont  [1]{#1}%
\providecommand \bibfnamefont [1]{#1}%
\providecommand \citenamefont [1]{#1}%
\providecommand \href@noop [0]{\@secondoftwo}%
\providecommand \href [0]{\begingroup \@sanitize@url \@href}%
\providecommand \@href[1]{\@@startlink{#1}\@@href}%
\providecommand \@@href[1]{\endgroup#1\@@endlink}%
\providecommand \@sanitize@url [0]{\catcode `\\12\catcode `\$12\catcode
  `\&12\catcode `\#12\catcode `\^12\catcode `\_12\catcode `\%12\relax}%
\providecommand \@@startlink[1]{}%
\providecommand \@@endlink[0]{}%
\providecommand \url  [0]{\begingroup\@sanitize@url \@url }%
\providecommand \@url [1]{\endgroup\@href {#1}{\urlprefix }}%
\providecommand \urlprefix  [0]{URL }%
\providecommand \Eprint [0]{\href }%
\providecommand \doibase [0]{http://dx.doi.org/}%
\providecommand \selectlanguage [0]{\@gobble}%
\providecommand \bibinfo  [0]{\@secondoftwo}%
\providecommand \bibfield  [0]{\@secondoftwo}%
\providecommand \translation [1]{[#1]}%
\providecommand \BibitemOpen [0]{}%
\providecommand \bibitemStop [0]{}%
\providecommand \bibitemNoStop [0]{.\EOS\space}%
\providecommand \EOS [0]{\spacefactor3000\relax}%
\providecommand \BibitemShut  [1]{\csname bibitem#1\endcsname}%
\let\auto@bib@innerbib\@empty
\bibitem [{\citenamefont {Wong}(1995)}]{Wong:1995jf}%
  \BibitemOpen
  \bibfield  {author} {\bibinfo {author} {\bibfnamefont {C.-Y.}\ \bibnamefont
  {Wong}},\ }\href {\doibase 10.1142/1128} {\emph {\bibinfo {title}
  {Introduction to High-Energy Heavy-Ion Collisions}}}\ (\bibinfo  {publisher}
  {WORLD SCIENTIFIC},\ \bibinfo {year} {1995})\BibitemShut {NoStop}%
\bibitem [{\citenamefont {Bertulani}\ and\ \citenamefont
  {Schechter}(2002)}]{Bertulani:2002kfo}%
  \BibitemOpen
  \bibfield  {author} {\bibinfo {author} {\bibfnamefont {C.~A.}\ \bibnamefont
  {Bertulani}}\ and\ \bibinfo {author} {\bibfnamefont {H.}~\bibnamefont
  {Schechter}},\ }\href@noop {} {\emph {\bibinfo {title} {{Introduction to
  Nuclear Physics}}}}\ (\bibinfo  {publisher} {Nova Science Publishers, Inc.},\
  \bibinfo {year} {2002})\BibitemShut {NoStop}%
\bibitem [{\citenamefont {Feynman}\ \emph {et~al.}(1978)\citenamefont
  {Feynman}, \citenamefont {Field},\ and\ \citenamefont
  {Fox}}]{Feynman:1978dt}%
  \BibitemOpen
  \bibfield  {author} {\bibinfo {author} {\bibfnamefont {R.~P.}\ \bibnamefont
  {Feynman}}, \bibinfo {author} {\bibfnamefont {R.~D.}\ \bibnamefont {Field}},
  \ and\ \bibinfo {author} {\bibfnamefont {G.~C.}\ \bibnamefont {Fox}},\ }\href
  {\doibase 10.1103/PhysRevD.18.3320} {\bibfield  {journal} {\bibinfo
  {journal} {Phys. Rev. D}\ }\textbf {\bibinfo {volume} {18}},\ \bibinfo
  {pages} {3320} (\bibinfo {year} {1978})}\BibitemShut {NoStop}%
\bibitem [{\citenamefont {Afanasiev}\ \emph {et~al.}(2002)\citenamefont
  {Afanasiev} \emph {et~al.}}]{NA49:2002pzu}%
  \BibitemOpen
  \bibfield  {author} {\bibinfo {author} {\bibfnamefont {S.~V.}\ \bibnamefont
  {Afanasiev}} \emph {et~al.} (\bibinfo {collaboration} {NA49}),\ }\href
  {\doibase 10.1103/PhysRevC.66.054902} {\bibfield  {journal} {\bibinfo
  {journal} {Phys. Rev. C}\ }\textbf {\bibinfo {volume} {66}},\ \bibinfo
  {pages} {054902} (\bibinfo {year} {2002})},\ \Eprint
  {http://arxiv.org/abs/nucl-ex/0205002} {arXiv:nucl-ex/0205002} \BibitemShut
  {NoStop}%
\bibitem [{\citenamefont {Alt}\ \emph {et~al.}(2005)\citenamefont {Alt} \emph
  {et~al.}}]{NA49:2004jzr}%
  \BibitemOpen
  \bibfield  {author} {\bibinfo {author} {\bibfnamefont {C.}~\bibnamefont
  {Alt}} \emph {et~al.} (\bibinfo {collaboration} {NA49}),\ }\href {\doibase
  10.1103/PhysRevLett.94.052301} {\bibfield  {journal} {\bibinfo  {journal}
  {Phys. Rev. Lett.}\ }\textbf {\bibinfo {volume} {94}},\ \bibinfo {pages}
  {052301} (\bibinfo {year} {2005})},\ \Eprint
  {http://arxiv.org/abs/nucl-ex/0406031} {arXiv:nucl-ex/0406031} \BibitemShut
  {NoStop}%
\bibitem [{\citenamefont {Alt}\ \emph {et~al.}(2006)\citenamefont {Alt} \emph
  {et~al.}}]{NA49:2006gaj}%
  \BibitemOpen
  \bibfield  {author} {\bibinfo {author} {\bibfnamefont {C.}~\bibnamefont
  {Alt}} \emph {et~al.} (\bibinfo {collaboration} {NA49}),\ }\href {\doibase
  10.1103/PhysRevC.73.044910} {\bibfield  {journal} {\bibinfo  {journal} {Phys.
  Rev. C}\ }\textbf {\bibinfo {volume} {73}},\ \bibinfo {pages} {044910}
  (\bibinfo {year} {2006})}\BibitemShut {NoStop}%
\bibitem [{\citenamefont {Alt}\ \emph {et~al.}(2008)\citenamefont {Alt} \emph
  {et~al.}}]{NA49:2007stj}%
  \BibitemOpen
  \bibfield  {author} {\bibinfo {author} {\bibfnamefont {C.}~\bibnamefont
  {Alt}} \emph {et~al.} (\bibinfo {collaboration} {NA49}),\ }\href {\doibase
  10.1103/PhysRevC.77.024903} {\bibfield  {journal} {\bibinfo  {journal} {Phys.
  Rev. C}\ }\textbf {\bibinfo {volume} {77}},\ \bibinfo {pages} {024903}
  (\bibinfo {year} {2008})},\ \Eprint {http://arxiv.org/abs/0710.0118}
  {arXiv:0710.0118 [nucl-ex]} \BibitemShut {NoStop}%
\bibitem [{\citenamefont {Adams}\ \emph {et~al.}(2004)\citenamefont {Adams}
  \emph {et~al.}}]{STAR:2003jwm}%
  \BibitemOpen
  \bibfield  {author} {\bibinfo {author} {\bibfnamefont {J.}~\bibnamefont
  {Adams}} \emph {et~al.} (\bibinfo {collaboration} {STAR}),\ }\href {\doibase
  10.1103/PhysRevLett.92.112301} {\bibfield  {journal} {\bibinfo  {journal}
  {Phys. Rev. Lett.}\ }\textbf {\bibinfo {volume} {92}},\ \bibinfo {pages}
  {112301} (\bibinfo {year} {2004})},\ \Eprint
  {http://arxiv.org/abs/nucl-ex/0310004} {arXiv:nucl-ex/0310004} \BibitemShut
  {NoStop}%
\bibitem [{\citenamefont {Abelev}\ \emph {et~al.}(2009)\citenamefont {Abelev}
  \emph {et~al.}}]{STAR:2008med}%
  \BibitemOpen
  \bibfield  {author} {\bibinfo {author} {\bibfnamefont {B.~I.}\ \bibnamefont
  {Abelev}} \emph {et~al.} (\bibinfo {collaboration} {STAR}),\ }\href {\doibase
  10.1103/PhysRevC.79.034909} {\bibfield  {journal} {\bibinfo  {journal} {Phys.
  Rev. C}\ }\textbf {\bibinfo {volume} {79}},\ \bibinfo {pages} {034909}
  (\bibinfo {year} {2009})},\ \Eprint {http://arxiv.org/abs/0808.2041}
  {arXiv:0808.2041 [nucl-ex]} \BibitemShut {NoStop}%
\bibitem [{\citenamefont {Agakishiev}\ \emph {et~al.}(2012)\citenamefont
  {Agakishiev} \emph {et~al.}}]{STAR:2011iap}%
  \BibitemOpen
  \bibfield  {author} {\bibinfo {author} {\bibfnamefont {G.}~\bibnamefont
  {Agakishiev}} \emph {et~al.} (\bibinfo {collaboration} {STAR}),\ }\href
  {\doibase 10.1103/PhysRevLett.108.072302} {\bibfield  {journal} {\bibinfo
  {journal} {Phys. Rev. Lett.}\ }\textbf {\bibinfo {volume} {108}},\ \bibinfo
  {pages} {072302} (\bibinfo {year} {2012})},\ \Eprint
  {http://arxiv.org/abs/1110.0579} {arXiv:1110.0579 [nucl-ex]} \BibitemShut
  {NoStop}%
\bibitem [{\citenamefont {Adamczyk}\ \emph {et~al.}(2017)\citenamefont
  {Adamczyk} \emph {et~al.}}]{STAR:2017sal}%
  \BibitemOpen
  \bibfield  {author} {\bibinfo {author} {\bibfnamefont {L.}~\bibnamefont
  {Adamczyk}} \emph {et~al.} (\bibinfo {collaboration} {STAR}),\ }\href
  {\doibase 10.1103/PhysRevC.96.044904} {\bibfield  {journal} {\bibinfo
  {journal} {Phys. Rev. C}\ }\textbf {\bibinfo {volume} {96}},\ \bibinfo
  {pages} {044904} (\bibinfo {year} {2017})},\ \Eprint
  {http://arxiv.org/abs/1701.07065} {arXiv:1701.07065 [nucl-ex]} \BibitemShut
  {NoStop}%
\bibitem [{\citenamefont {Abelev}\ \emph
  {et~al.}(2013{\natexlab{a}})\citenamefont {Abelev} \emph
  {et~al.}}]{ALICE:2013mez}%
  \BibitemOpen
  \bibfield  {author} {\bibinfo {author} {\bibfnamefont {B.}~\bibnamefont
  {Abelev}} \emph {et~al.} (\bibinfo {collaboration} {ALICE}),\ }\href
  {\doibase 10.1103/PhysRevC.88.044910} {\bibfield  {journal} {\bibinfo
  {journal} {Phys. Rev. C}\ }\textbf {\bibinfo {volume} {88}},\ \bibinfo
  {pages} {044910} (\bibinfo {year} {2013}{\natexlab{a}})},\ \Eprint
  {http://arxiv.org/abs/1303.0737} {arXiv:1303.0737 [hep-ex]} \BibitemShut
  {NoStop}%
\bibitem [{\citenamefont {Abelev}\ \emph
  {et~al.}(2014{\natexlab{a}})\citenamefont {Abelev} \emph
  {et~al.}}]{ALICE:2013wgn}%
  \BibitemOpen
  \bibfield  {author} {\bibinfo {author} {\bibfnamefont {B.~B.}\ \bibnamefont
  {Abelev}} \emph {et~al.} (\bibinfo {collaboration} {ALICE}),\ }\href
  {\doibase 10.1016/j.physletb.2013.11.020} {\bibfield  {journal} {\bibinfo
  {journal} {Phys. Lett. B}\ }\textbf {\bibinfo {volume} {728}},\ \bibinfo
  {pages} {25} (\bibinfo {year} {2014}{\natexlab{a}})},\ \Eprint
  {http://arxiv.org/abs/1307.6796} {arXiv:1307.6796 [nucl-ex]} \BibitemShut
  {NoStop}%
\bibitem [{\citenamefont {de~Florian}\ \emph {et~al.}(2017)\citenamefont
  {de~Florian}, \citenamefont {Epele}, \citenamefont {Hernandez-Pinto},
  \citenamefont {Sassot},\ and\ \citenamefont {Stratmann}}]{deFlorian:2017lwf}%
  \BibitemOpen
  \bibfield  {author} {\bibinfo {author} {\bibfnamefont {D.}~\bibnamefont
  {de~Florian}}, \bibinfo {author} {\bibfnamefont {M.}~\bibnamefont {Epele}},
  \bibinfo {author} {\bibfnamefont {R.~J.}\ \bibnamefont {Hernandez-Pinto}},
  \bibinfo {author} {\bibfnamefont {R.}~\bibnamefont {Sassot}}, \ and\ \bibinfo
  {author} {\bibfnamefont {M.}~\bibnamefont {Stratmann}},\ }\href {\doibase
  10.1103/PhysRevD.95.094019} {\bibfield  {journal} {\bibinfo  {journal} {Phys.
  Rev. D}\ }\textbf {\bibinfo {volume} {95}},\ \bibinfo {pages} {094019}
  (\bibinfo {year} {2017})},\ \Eprint {http://arxiv.org/abs/1702.06353}
  {arXiv:1702.06353 [hep-ph]} \BibitemShut {NoStop}%
\bibitem [{\citenamefont {Albino}\ \emph {et~al.}(2008)\citenamefont {Albino},
  \citenamefont {Kniehl},\ and\ \citenamefont {Kramer}}]{Albino:2008fy}%
  \BibitemOpen
  \bibfield  {author} {\bibinfo {author} {\bibfnamefont {S.}~\bibnamefont
  {Albino}}, \bibinfo {author} {\bibfnamefont {B.~A.}\ \bibnamefont {Kniehl}},
  \ and\ \bibinfo {author} {\bibfnamefont {G.}~\bibnamefont {Kramer}},\ }\href
  {\doibase 10.1016/j.nuclphysb.2008.05.017} {\bibfield  {journal} {\bibinfo
  {journal} {Nucl. Phys. B}\ }\textbf {\bibinfo {volume} {803}},\ \bibinfo
  {pages} {42} (\bibinfo {year} {2008})},\ \Eprint
  {http://arxiv.org/abs/0803.2768} {arXiv:0803.2768 [hep-ph]} \BibitemShut
  {NoStop}%
\bibitem [{\citenamefont {de~Florian}\ \emph {et~al.}(2007)\citenamefont
  {de~Florian}, \citenamefont {Sassot},\ and\ \citenamefont
  {Stratmann}}]{deFlorian:2007ekg}%
  \BibitemOpen
  \bibfield  {author} {\bibinfo {author} {\bibfnamefont {D.}~\bibnamefont
  {de~Florian}}, \bibinfo {author} {\bibfnamefont {R.}~\bibnamefont {Sassot}},
  \ and\ \bibinfo {author} {\bibfnamefont {M.}~\bibnamefont {Stratmann}},\
  }\href {\doibase 10.1103/PhysRevD.76.074033} {\bibfield  {journal} {\bibinfo
  {journal} {Phys. Rev. D}\ }\textbf {\bibinfo {volume} {76}},\ \bibinfo
  {pages} {074033} (\bibinfo {year} {2007})},\ \Eprint
  {http://arxiv.org/abs/0707.1506} {arXiv:0707.1506 [hep-ph]} \BibitemShut
  {NoStop}%
\bibitem [{\citenamefont {Hirano}\ and\ \citenamefont
  {Nara}(2004)}]{Hirano:2003pw}%
  \BibitemOpen
  \bibfield  {author} {\bibinfo {author} {\bibfnamefont {T.}~\bibnamefont
  {Hirano}}\ and\ \bibinfo {author} {\bibfnamefont {Y.}~\bibnamefont {Nara}},\
  }\href {\doibase 10.1103/PhysRevC.69.034908} {\bibfield  {journal} {\bibinfo
  {journal} {Phys. Rev. C}\ }\textbf {\bibinfo {volume} {69}},\ \bibinfo
  {pages} {034908} (\bibinfo {year} {2004})},\ \Eprint
  {http://arxiv.org/abs/nucl-th/0307015} {arXiv:nucl-th/0307015} \BibitemShut
  {NoStop}%
\bibitem [{\citenamefont {Eskola}\ \emph {et~al.}(2005)\citenamefont {Eskola},
  \citenamefont {Honkanen}, \citenamefont {Niemi}, \citenamefont {Ruuskanen},\
  and\ \citenamefont {Rasanen}}]{Eskola:2005ue}%
  \BibitemOpen
  \bibfield  {author} {\bibinfo {author} {\bibfnamefont {K.~J.}\ \bibnamefont
  {Eskola}}, \bibinfo {author} {\bibfnamefont {H.}~\bibnamefont {Honkanen}},
  \bibinfo {author} {\bibfnamefont {H.}~\bibnamefont {Niemi}}, \bibinfo
  {author} {\bibfnamefont {P.~V.}\ \bibnamefont {Ruuskanen}}, \ and\ \bibinfo
  {author} {\bibfnamefont {S.~S.}\ \bibnamefont {Rasanen}},\ }\href {\doibase
  10.1103/PhysRevC.72.044904} {\bibfield  {journal} {\bibinfo  {journal} {Phys.
  Rev. C}\ }\textbf {\bibinfo {volume} {72}},\ \bibinfo {pages} {044904}
  (\bibinfo {year} {2005})},\ \Eprint {http://arxiv.org/abs/hep-ph/0506049}
  {arXiv:hep-ph/0506049} \BibitemShut {NoStop}%
\bibitem [{\citenamefont {Tsallis}(1988)}]{Tsallis:1987eu}%
  \BibitemOpen
  \bibfield  {author} {\bibinfo {author} {\bibfnamefont {C.}~\bibnamefont
  {Tsallis}},\ }\href {\doibase 10.1007/BF01016429} {\bibfield  {journal}
  {\bibinfo  {journal} {J. Statist. Phys.}\ }\textbf {\bibinfo {volume} {52}},\
  \bibinfo {pages} {479} (\bibinfo {year} {1988})}\BibitemShut {NoStop}%
\bibitem [{\citenamefont {Abelev}\ \emph
  {et~al.}(2014{\natexlab{b}})\citenamefont {Abelev} \emph
  {et~al.}}]{ALICE:2014juv}%
  \BibitemOpen
  \bibfield  {author} {\bibinfo {author} {\bibfnamefont {B.~B.}\ \bibnamefont
  {Abelev}} \emph {et~al.} (\bibinfo {collaboration} {ALICE}),\ }\href
  {\doibase 10.1016/j.physletb.2014.07.011} {\bibfield  {journal} {\bibinfo
  {journal} {Phys. Lett. B}\ }\textbf {\bibinfo {volume} {736}},\ \bibinfo
  {pages} {196} (\bibinfo {year} {2014}{\natexlab{b}})},\ \Eprint
  {http://arxiv.org/abs/1401.1250} {arXiv:1401.1250 [nucl-ex]} \BibitemShut
  {NoStop}%
\bibitem [{\citenamefont {Abelev}\ \emph
  {et~al.}(2013{\natexlab{b}})\citenamefont {Abelev} \emph
  {et~al.}}]{ALICE:2013cdo}%
  \BibitemOpen
  \bibfield  {author} {\bibinfo {author} {\bibfnamefont {B.~B.}\ \bibnamefont
  {Abelev}} \emph {et~al.} (\bibinfo {collaboration} {ALICE}),\ }\href
  {\doibase 10.1103/PhysRevLett.111.222301} {\bibfield  {journal} {\bibinfo
  {journal} {Phys. Rev. Lett.}\ }\textbf {\bibinfo {volume} {111}},\ \bibinfo
  {pages} {222301} (\bibinfo {year} {2013}{\natexlab{b}})},\ \Eprint
  {http://arxiv.org/abs/1307.5530} {arXiv:1307.5530 [nucl-ex]} \BibitemShut
  {NoStop}%
\bibitem [{\citenamefont {Acharya}\ \emph {et~al.}(2021)\citenamefont {Acharya}
  \emph {et~al.}}]{ALICE:2020jsh}%
  \BibitemOpen
  \bibfield  {author} {\bibinfo {author} {\bibfnamefont {S.}~\bibnamefont
  {Acharya}} \emph {et~al.} (\bibinfo {collaboration} {ALICE}),\ }\href
  {\doibase 10.1140/epjc/s10052-020-08690-5} {\bibfield  {journal} {\bibinfo
  {journal} {Eur. Phys. J. C}\ }\textbf {\bibinfo {volume} {81}},\ \bibinfo
  {pages} {256} (\bibinfo {year} {2021})},\ \Eprint
  {http://arxiv.org/abs/2005.11120} {arXiv:2005.11120 [nucl-ex]} \BibitemShut
  {NoStop}%
\bibitem [{\citenamefont {Acharya}\ \emph {et~al.}(2024)\citenamefont {Acharya}
  \emph {et~al.}}]{ALICE:2023egx}%
  \BibitemOpen
  \bibfield  {author} {\bibinfo {author} {\bibfnamefont {S.}~\bibnamefont
  {Acharya}} \emph {et~al.} (\bibinfo {collaboration} {ALICE}),\ }\href
  {\doibase 10.1007/JHEP05(2024)317} {\bibfield  {journal} {\bibinfo  {journal}
  {JHEP}\ }\textbf {\bibinfo {volume} {05}},\ \bibinfo {pages} {317} (\bibinfo
  {year} {2024})},\ \Eprint {http://arxiv.org/abs/2308.16116} {arXiv:2308.16116
  [nucl-ex]} \BibitemShut {NoStop}%
\bibitem [{\citenamefont {Acharya}\ \emph {et~al.}(2022)\citenamefont {Acharya}
  \emph {et~al.}}]{ALICE:2021ptz}%
  \BibitemOpen
  \bibfield  {author} {\bibinfo {author} {\bibfnamefont {S.}~\bibnamefont
  {Acharya}} \emph {et~al.} (\bibinfo {collaboration} {ALICE}),\ }\href
  {\doibase 10.1103/PhysRevC.106.034907} {\bibfield  {journal} {\bibinfo
  {journal} {Phys. Rev. C}\ }\textbf {\bibinfo {volume} {106}},\ \bibinfo
  {pages} {034907} (\bibinfo {year} {2022})},\ \Eprint
  {http://arxiv.org/abs/2106.13113} {arXiv:2106.13113 [nucl-ex]} \BibitemShut
  {NoStop}%
\bibitem [{\citenamefont {Acharya}\ \emph
  {et~al.}(2020{\natexlab{a}})\citenamefont {Acharya} \emph
  {et~al.}}]{ALICE:2019hno}%
  \BibitemOpen
  \bibfield  {author} {\bibinfo {author} {\bibfnamefont {S.}~\bibnamefont
  {Acharya}} \emph {et~al.} (\bibinfo {collaboration} {ALICE}),\ }\href
  {\doibase 10.1103/PhysRevC.101.044907} {\bibfield  {journal} {\bibinfo
  {journal} {Phys. Rev. C}\ }\textbf {\bibinfo {volume} {101}},\ \bibinfo
  {pages} {044907} (\bibinfo {year} {2020}{\natexlab{a}})},\ \Eprint
  {http://arxiv.org/abs/1910.07678} {arXiv:1910.07678 [nucl-ex]} \BibitemShut
  {NoStop}%
\bibitem [{\citenamefont {Adam}\ \emph
  {et~al.}(2017{\natexlab{a}})\citenamefont {Adam} \emph
  {et~al.}}]{ALICE:2017ban}%
  \BibitemOpen
  \bibfield  {author} {\bibinfo {author} {\bibfnamefont {J.}~\bibnamefont
  {Adam}} \emph {et~al.} (\bibinfo {collaboration} {ALICE}),\ }\href {\doibase
  10.1103/PhysRevC.95.064606} {\bibfield  {journal} {\bibinfo  {journal} {Phys.
  Rev. C}\ }\textbf {\bibinfo {volume} {95}},\ \bibinfo {pages} {064606}
  (\bibinfo {year} {2017}{\natexlab{a}})},\ \Eprint
  {http://arxiv.org/abs/1702.00555} {arXiv:1702.00555 [nucl-ex]} \BibitemShut
  {NoStop}%
\bibitem [{\citenamefont {Abelev}\ \emph {et~al.}(2007)\citenamefont {Abelev}
  \emph {et~al.}}]{STAR:2006nmo}%
  \BibitemOpen
  \bibfield  {author} {\bibinfo {author} {\bibfnamefont {B.~I.}\ \bibnamefont
  {Abelev}} \emph {et~al.} (\bibinfo {collaboration} {STAR}),\ }\href {\doibase
  10.1103/PhysRevC.75.064901} {\bibfield  {journal} {\bibinfo  {journal} {Phys.
  Rev. C}\ }\textbf {\bibinfo {volume} {75}},\ \bibinfo {pages} {064901}
  (\bibinfo {year} {2007})},\ \Eprint {http://arxiv.org/abs/nucl-ex/0607033}
  {arXiv:nucl-ex/0607033} \BibitemShut {NoStop}%
\bibitem [{\citenamefont {Adams}\ \emph {et~al.}(2006)\citenamefont {Adams}
  \emph {et~al.}}]{STAR:2006xud}%
  \BibitemOpen
  \bibfield  {author} {\bibinfo {author} {\bibfnamefont {J.}~\bibnamefont
  {Adams}} \emph {et~al.} (\bibinfo {collaboration} {STAR}),\ }\href {\doibase
  10.1016/j.physletb.2006.04.032} {\bibfield  {journal} {\bibinfo  {journal}
  {Phys. Lett. B}\ }\textbf {\bibinfo {volume} {637}},\ \bibinfo {pages} {161}
  (\bibinfo {year} {2006})},\ \Eprint {http://arxiv.org/abs/nucl-ex/0601033}
  {arXiv:nucl-ex/0601033} \BibitemShut {NoStop}%
\bibitem [{\citenamefont {Khachatryan}\ \emph {et~al.}(2011)\citenamefont
  {Khachatryan} \emph {et~al.}}]{CMS:2011jlm}%
  \BibitemOpen
  \bibfield  {author} {\bibinfo {author} {\bibfnamefont {V.}~\bibnamefont
  {Khachatryan}} \emph {et~al.} (\bibinfo {collaboration} {CMS}),\ }\href
  {\doibase 10.1007/JHEP05(2011)064} {\bibfield  {journal} {\bibinfo  {journal}
  {JHEP}\ }\textbf {\bibinfo {volume} {05}},\ \bibinfo {pages} {064} (\bibinfo
  {year} {2011})},\ \Eprint {http://arxiv.org/abs/1102.4282} {arXiv:1102.4282
  [hep-ex]} \BibitemShut {NoStop}%
\bibitem [{\citenamefont {Acharya}\ \emph
  {et~al.}(2020{\natexlab{b}})\citenamefont {Acharya} \emph
  {et~al.}}]{ALICE:2020nkc}%
  \BibitemOpen
  \bibfield  {author} {\bibinfo {author} {\bibfnamefont {S.}~\bibnamefont
  {Acharya}} \emph {et~al.} (\bibinfo {collaboration} {ALICE}),\ }\href
  {\doibase 10.1140/epjc/s10052-020-8125-1} {\bibfield  {journal} {\bibinfo
  {journal} {Eur. Phys. J. C}\ }\textbf {\bibinfo {volume} {80}},\ \bibinfo
  {pages} {693} (\bibinfo {year} {2020}{\natexlab{b}})},\ \Eprint
  {http://arxiv.org/abs/2003.02394} {arXiv:2003.02394 [nucl-ex]} \BibitemShut
  {NoStop}%
\bibitem [{\citenamefont {Acharya}\ \emph
  {et~al.}(2020{\natexlab{c}})\citenamefont {Acharya} \emph
  {et~al.}}]{ALICE:2019etb}%
  \BibitemOpen
  \bibfield  {author} {\bibinfo {author} {\bibfnamefont {S.}~\bibnamefont
  {Acharya}} \emph {et~al.} (\bibinfo {collaboration} {ALICE}),\ }\href
  {\doibase 10.1016/j.physletb.2020.135501} {\bibfield  {journal} {\bibinfo
  {journal} {Phys. Lett. B}\ }\textbf {\bibinfo {volume} {807}},\ \bibinfo
  {pages} {135501} (\bibinfo {year} {2020}{\natexlab{c}})},\ \Eprint
  {http://arxiv.org/abs/1910.14397} {arXiv:1910.14397 [nucl-ex]} \BibitemShut
  {NoStop}%
\bibitem [{\citenamefont {Acharya}\ \emph
  {et~al.}(2020{\natexlab{d}})\citenamefont {Acharya} \emph
  {et~al.}}]{ALICE:2019avo}%
  \BibitemOpen
  \bibfield  {author} {\bibinfo {author} {\bibfnamefont {S.}~\bibnamefont
  {Acharya}} \emph {et~al.} (\bibinfo {collaboration} {ALICE}),\ }\href
  {\doibase 10.1140/epjc/s10052-020-7673-8} {\bibfield  {journal} {\bibinfo
  {journal} {Eur. Phys. J. C}\ }\textbf {\bibinfo {volume} {80}},\ \bibinfo
  {pages} {167} (\bibinfo {year} {2020}{\natexlab{d}})},\ \Eprint
  {http://arxiv.org/abs/1908.01861} {arXiv:1908.01861 [nucl-ex]} \BibitemShut
  {NoStop}%
\bibitem [{\citenamefont {Adam}\ \emph
  {et~al.}(2017{\natexlab{b}})\citenamefont {Adam} \emph
  {et~al.}}]{ALICE:2016fzo}%
  \BibitemOpen
  \bibfield  {author} {\bibinfo {author} {\bibfnamefont {J.}~\bibnamefont
  {Adam}} \emph {et~al.} (\bibinfo {collaboration} {ALICE}),\ }\href {\doibase
  10.1038/nphys4111} {\bibfield  {journal} {\bibinfo  {journal} {Nature Phys.}\
  }\textbf {\bibinfo {volume} {13}},\ \bibinfo {pages} {535} (\bibinfo {year}
  {2017}{\natexlab{b}})},\ \Eprint {http://arxiv.org/abs/1606.07424}
  {arXiv:1606.07424 [nucl-ex]} \BibitemShut {NoStop}%
\bibitem [{\citenamefont {Acharya}\ \emph {et~al.}(2019)\citenamefont {Acharya}
  \emph {et~al.}}]{ALICE:2018pal}%
  \BibitemOpen
  \bibfield  {author} {\bibinfo {author} {\bibfnamefont {S.}~\bibnamefont
  {Acharya}} \emph {et~al.} (\bibinfo {collaboration} {ALICE}),\ }\href
  {\doibase 10.1103/PhysRevC.99.024906} {\bibfield  {journal} {\bibinfo
  {journal} {Phys. Rev. C}\ }\textbf {\bibinfo {volume} {99}},\ \bibinfo
  {pages} {024906} (\bibinfo {year} {2019})},\ \Eprint
  {http://arxiv.org/abs/1807.11321} {arXiv:1807.11321 [nucl-ex]} \BibitemShut
  {NoStop}%
\bibitem [{\citenamefont {Adam}\ \emph
  {et~al.}(2016{\natexlab{a}})\citenamefont {Adam} \emph
  {et~al.}}]{ALICE:2016dei}%
  \BibitemOpen
  \bibfield  {author} {\bibinfo {author} {\bibfnamefont {J.}~\bibnamefont
  {Adam}} \emph {et~al.} (\bibinfo {collaboration} {ALICE}),\ }\href {\doibase
  10.1016/j.physletb.2016.07.050} {\bibfield  {journal} {\bibinfo  {journal}
  {Phys. Lett. B}\ }\textbf {\bibinfo {volume} {760}},\ \bibinfo {pages} {720}
  (\bibinfo {year} {2016}{\natexlab{a}})},\ \Eprint
  {http://arxiv.org/abs/1601.03658} {arXiv:1601.03658 [nucl-ex]} \BibitemShut
  {NoStop}%
\bibitem [{\citenamefont {Adam}\ \emph
  {et~al.}(2016{\natexlab{b}})\citenamefont {Adam} \emph
  {et~al.}}]{ALICE:2016sak}%
  \BibitemOpen
  \bibfield  {author} {\bibinfo {author} {\bibfnamefont {J.}~\bibnamefont
  {Adam}} \emph {et~al.} (\bibinfo {collaboration} {ALICE}),\ }\href {\doibase
  10.1140/epjc/s10052-016-4088-7} {\bibfield  {journal} {\bibinfo  {journal}
  {Eur. Phys. J. C}\ }\textbf {\bibinfo {volume} {76}},\ \bibinfo {pages} {245}
  (\bibinfo {year} {2016}{\natexlab{b}})},\ \Eprint
  {http://arxiv.org/abs/1601.07868} {arXiv:1601.07868 [nucl-ex]} \BibitemShut
  {NoStop}%
\bibitem [{\citenamefont {Adam}\ \emph {et~al.}(2020)\citenamefont {Adam} \emph
  {et~al.}}]{STAR:2019bjj}%
  \BibitemOpen
  \bibfield  {author} {\bibinfo {author} {\bibfnamefont {J.}~\bibnamefont
  {Adam}} \emph {et~al.} (\bibinfo {collaboration} {STAR}),\ }\href {\doibase
  10.1103/PhysRevC.102.034909} {\bibfield  {journal} {\bibinfo  {journal}
  {Phys. Rev. C}\ }\textbf {\bibinfo {volume} {102}},\ \bibinfo {pages}
  {034909} (\bibinfo {year} {2020})},\ \Eprint
  {http://arxiv.org/abs/1906.03732} {arXiv:1906.03732 [nucl-ex]} \BibitemShut
  {NoStop}%
\bibitem [{\citenamefont {Acharya}\ \emph
  {et~al.}(2020{\natexlab{e}})\citenamefont {Acharya} \emph
  {et~al.}}]{ALICE:2019xyr}%
  \BibitemOpen
  \bibfield  {author} {\bibinfo {author} {\bibfnamefont {S.}~\bibnamefont
  {Acharya}} \emph {et~al.} (\bibinfo {collaboration} {ALICE}),\ }\href
  {\doibase 10.1016/j.physletb.2020.135225} {\bibfield  {journal} {\bibinfo
  {journal} {Phys. Lett. B}\ }\textbf {\bibinfo {volume} {802}},\ \bibinfo
  {pages} {135225} (\bibinfo {year} {2020}{\natexlab{e}})},\ \Eprint
  {http://arxiv.org/abs/1910.14419} {arXiv:1910.14419 [nucl-ex]} \BibitemShut
  {NoStop}%
\bibitem [{\citenamefont {Song}\ \emph {et~al.}(2017)\citenamefont {Song},
  \citenamefont {Gou}, \citenamefont {Shao},\ and\ \citenamefont
  {Liang}}]{Song:2017gcz}%
  \BibitemOpen
  \bibfield  {author} {\bibinfo {author} {\bibfnamefont {J.}~\bibnamefont
  {Song}}, \bibinfo {author} {\bibfnamefont {X.-r.}\ \bibnamefont {Gou}},
  \bibinfo {author} {\bibfnamefont {F.-l.}\ \bibnamefont {Shao}}, \ and\
  \bibinfo {author} {\bibfnamefont {Z.-T.}\ \bibnamefont {Liang}},\ }\href
  {\doibase 10.1016/j.physletb.2017.10.012} {\bibfield  {journal} {\bibinfo
  {journal} {Phys. Lett.}\ }\textbf {\bibinfo {volume} {B774}},\ \bibinfo
  {pages} {516} (\bibinfo {year} {2017})},\ \Eprint
  {http://arxiv.org/abs/1707.03949} {arXiv:1707.03949 [hep-ph]} \BibitemShut
  {NoStop}%
\bibitem [{\citenamefont {Gou}\ \emph {et~al.}(2017)\citenamefont {Gou},
  \citenamefont {Shao}, \citenamefont {Wang}, \citenamefont {Li},\ and\
  \citenamefont {Song}}]{Gou:2017foe}%
  \BibitemOpen
  \bibfield  {author} {\bibinfo {author} {\bibfnamefont {X.-r.}\ \bibnamefont
  {Gou}}, \bibinfo {author} {\bibfnamefont {F.-l.}\ \bibnamefont {Shao}},
  \bibinfo {author} {\bibfnamefont {R.-q.}\ \bibnamefont {Wang}}, \bibinfo
  {author} {\bibfnamefont {H.-h.}\ \bibnamefont {Li}}, \ and\ \bibinfo {author}
  {\bibfnamefont {J.}~\bibnamefont {Song}},\ }\href {\doibase
  10.1103/PhysRevD.96.094010} {\bibfield  {journal} {\bibinfo  {journal} {Phys.
  Rev.}\ }\textbf {\bibinfo {volume} {D96}},\ \bibinfo {pages} {094010}
  (\bibinfo {year} {2017})},\ \Eprint {http://arxiv.org/abs/1707.06906}
  {arXiv:1707.06906 [hep-ph]} \BibitemShut {NoStop}%
\bibitem [{\citenamefont {Li}\ \emph {et~al.}(2021)\citenamefont {Li},
  \citenamefont {Shao},\ and\ \citenamefont {Song}}]{Li:2021nhq}%
  \BibitemOpen
  \bibfield  {author} {\bibinfo {author} {\bibfnamefont {H.-h.}\ \bibnamefont
  {Li}}, \bibinfo {author} {\bibfnamefont {F.-l.}\ \bibnamefont {Shao}}, \ and\
  \bibinfo {author} {\bibfnamefont {J.}~\bibnamefont {Song}},\ }\href {\doibase
  10.1088/1674-1137/ac1ef9} {\bibfield  {journal} {\bibinfo  {journal} {Chin.
  Phys. C}\ }\textbf {\bibinfo {volume} {45}},\ \bibinfo {pages} {113105}
  (\bibinfo {year} {2021})},\ \Eprint {http://arxiv.org/abs/2103.14900}
  {arXiv:2103.14900 [hep-ph]} \BibitemShut {NoStop}%
\bibitem [{\citenamefont {Song}\ \emph {et~al.}(2021)\citenamefont {Song},
  \citenamefont {Wang}, \citenamefont {Li}, \citenamefont {Wang},\ and\
  \citenamefont {Shao}}]{Song:2020kak}%
  \BibitemOpen
  \bibfield  {author} {\bibinfo {author} {\bibfnamefont {J.}~\bibnamefont
  {Song}}, \bibinfo {author} {\bibfnamefont {X.-F.}\ \bibnamefont {Wang}},
  \bibinfo {author} {\bibfnamefont {H.-H.}\ \bibnamefont {Li}}, \bibinfo
  {author} {\bibfnamefont {R.-Q.}\ \bibnamefont {Wang}}, \ and\ \bibinfo
  {author} {\bibfnamefont {F.-L.}\ \bibnamefont {Shao}},\ }\href {\doibase
  10.1103/PhysRevC.103.034907} {\bibfield  {journal} {\bibinfo  {journal}
  {Phys. Rev. C}\ }\textbf {\bibinfo {volume} {103}},\ \bibinfo {pages}
  {034907} (\bibinfo {year} {2021})},\ \Eprint
  {http://arxiv.org/abs/2007.14588} {arXiv:2007.14588 [nucl-th]} \BibitemShut
  {NoStop}%
\bibitem [{\citenamefont {Song}\ \emph {et~al.}(2020)\citenamefont {Song},
  \citenamefont {Shao},\ and\ \citenamefont {Liang}}]{Song:2019sez}%
  \BibitemOpen
  \bibfield  {author} {\bibinfo {author} {\bibfnamefont {J.}~\bibnamefont
  {Song}}, \bibinfo {author} {\bibfnamefont {F.-l.}\ \bibnamefont {Shao}}, \
  and\ \bibinfo {author} {\bibfnamefont {Z.-t.}\ \bibnamefont {Liang}},\ }\href
  {\doibase 10.1103/PhysRevC.102.014911} {\bibfield  {journal} {\bibinfo
  {journal} {Phys. Rev. C}\ }\textbf {\bibinfo {volume} {102}},\ \bibinfo
  {pages} {014911} (\bibinfo {year} {2020})},\ \Eprint
  {http://arxiv.org/abs/1911.01152} {arXiv:1911.01152 [nucl-th]} \BibitemShut
  {NoStop}%
\bibitem [{\citenamefont {Li}\ \emph {et~al.}(2018)\citenamefont {Li},
  \citenamefont {Shao}, \citenamefont {Song},\ and\ \citenamefont
  {Wang}}]{Li:2017zuj}%
  \BibitemOpen
  \bibfield  {author} {\bibinfo {author} {\bibfnamefont {H.-H.}\ \bibnamefont
  {Li}}, \bibinfo {author} {\bibfnamefont {F.-L.}\ \bibnamefont {Shao}},
  \bibinfo {author} {\bibfnamefont {J.}~\bibnamefont {Song}}, \ and\ \bibinfo
  {author} {\bibfnamefont {R.-Q.}\ \bibnamefont {Wang}},\ }\href {\doibase
  10.1103/PhysRevC.97.064915} {\bibfield  {journal} {\bibinfo  {journal} {Phys.
  Rev.}\ }\textbf {\bibinfo {volume} {C97}},\ \bibinfo {pages} {064915}
  (\bibinfo {year} {2018})},\ \Eprint {http://arxiv.org/abs/1712.08921}
  {arXiv:1712.08921 [hep-ph]} \BibitemShut {NoStop}%
\bibitem [{\citenamefont {Song}\ \emph {et~al.}(2018)\citenamefont {Song},
  \citenamefont {Li},\ and\ \citenamefont {Shao}}]{Song:2018tpv}%
  \BibitemOpen
  \bibfield  {author} {\bibinfo {author} {\bibfnamefont {J.}~\bibnamefont
  {Song}}, \bibinfo {author} {\bibfnamefont {H.-h.}\ \bibnamefont {Li}}, \ and\
  \bibinfo {author} {\bibfnamefont {F.-l.}\ \bibnamefont {Shao}},\ }\href
  {\doibase 10.1140/epjc/s10052-018-5817-x} {\bibfield  {journal} {\bibinfo
  {journal} {Eur. Phys. J.}\ }\textbf {\bibinfo {volume} {C78}},\ \bibinfo
  {pages} {344} (\bibinfo {year} {2018})},\ \Eprint
  {http://arxiv.org/abs/1801.09402} {arXiv:1801.09402 [hep-ph]} \BibitemShut
  {NoStop}%
\bibitem [{\citenamefont {Song}\ \emph {et~al.}(2022)\citenamefont {Song},
  \citenamefont {Li},\ and\ \citenamefont {Shao}}]{Song:2021ojn}%
  \BibitemOpen
  \bibfield  {author} {\bibinfo {author} {\bibfnamefont {J.}~\bibnamefont
  {Song}}, \bibinfo {author} {\bibfnamefont {H.-h.}\ \bibnamefont {Li}}, \ and\
  \bibinfo {author} {\bibfnamefont {F.-l.}\ \bibnamefont {Shao}},\ }\href
  {\doibase 10.1103/PhysRevD.105.074027} {\bibfield  {journal} {\bibinfo
  {journal} {Phys. Rev. D}\ }\textbf {\bibinfo {volume} {105}},\ \bibinfo
  {pages} {074027} (\bibinfo {year} {2022})},\ \Eprint
  {http://arxiv.org/abs/2109.11722} {arXiv:2109.11722 [hep-ph]} \BibitemShut
  {NoStop}%
\bibitem [{\citenamefont {Song}\ \emph {et~al.}(2023)\citenamefont {Song},
  \citenamefont {Li},\ and\ \citenamefont {Shao}}]{Song:2023nzu}%
  \BibitemOpen
  \bibfield  {author} {\bibinfo {author} {\bibfnamefont {J.}~\bibnamefont
  {Song}}, \bibinfo {author} {\bibfnamefont {H.-h.}\ \bibnamefont {Li}}, \ and\
  \bibinfo {author} {\bibfnamefont {F.-l.}\ \bibnamefont {Shao}},\ }\href
  {\doibase 10.1140/epjc/s10052-023-12007-7} {\bibfield  {journal} {\bibinfo
  {journal} {Eur. Phys. J. C}\ }\textbf {\bibinfo {volume} {83}},\ \bibinfo
  {pages} {852} (\bibinfo {year} {2023})},\ \Eprint
  {http://arxiv.org/abs/2304.00434} {arXiv:2304.00434 [hep-ph]} \BibitemShut
  {NoStop}%
\bibitem [{\citenamefont {Alemany}\ and\ \citenamefont
  {Zanette}(1994)}]{Alemany:1994a}%
  \BibitemOpen
  \bibfield  {author} {\bibinfo {author} {\bibfnamefont {P.~A.}\ \bibnamefont
  {Alemany}}\ and\ \bibinfo {author} {\bibfnamefont {D.~H.}\ \bibnamefont
  {Zanette}},\ }\href {\doibase 10.1103/PhysRevE.49.R956} {\bibfield  {journal}
  {\bibinfo  {journal} {Phys. Rev. E}\ }\textbf {\bibinfo {volume} {49}},\
  \bibinfo {pages} {R956} (\bibinfo {year} {1994})}\BibitemShut {NoStop}%
\bibitem [{\citenamefont {Huovinen}\ and\ \citenamefont
  {Ruuskanen}(2006)}]{Huovinen:2006jp}%
  \BibitemOpen
  \bibfield  {author} {\bibinfo {author} {\bibfnamefont {P.}~\bibnamefont
  {Huovinen}}\ and\ \bibinfo {author} {\bibfnamefont {P.~V.}\ \bibnamefont
  {Ruuskanen}},\ }\href {\doibase 10.1146/annurev.nucl.54.070103.181236}
  {\bibfield  {journal} {\bibinfo  {journal} {Ann. Rev. Nucl. Part. Sci.}\
  }\textbf {\bibinfo {volume} {56}},\ \bibinfo {pages} {163} (\bibinfo {year}
  {2006})},\ \Eprint {http://arxiv.org/abs/nucl-th/0605008}
  {arXiv:nucl-th/0605008} \BibitemShut {NoStop}%
\bibitem [{\citenamefont {Kestin}\ and\ \citenamefont
  {Heinz}(2009)}]{Kestin:2008bh}%
  \BibitemOpen
  \bibfield  {author} {\bibinfo {author} {\bibfnamefont {G.}~\bibnamefont
  {Kestin}}\ and\ \bibinfo {author} {\bibfnamefont {U.~W.}\ \bibnamefont
  {Heinz}},\ }\href {\doibase 10.1140/epjc/s10052-008-0832-y} {\bibfield
  {journal} {\bibinfo  {journal} {Eur. Phys. J. C}\ }\textbf {\bibinfo {volume}
  {61}},\ \bibinfo {pages} {545} (\bibinfo {year} {2009})},\ \Eprint
  {http://arxiv.org/abs/0806.4539} {arXiv:0806.4539 [nucl-th]} \BibitemShut
  {NoStop}%
\bibitem [{\citenamefont {Song}\ \emph {et~al.}(2014)\citenamefont {Song},
  \citenamefont {Bass},\ and\ \citenamefont {Heinz}}]{Song:2013qma}%
  \BibitemOpen
  \bibfield  {author} {\bibinfo {author} {\bibfnamefont {H.}~\bibnamefont
  {Song}}, \bibinfo {author} {\bibfnamefont {S.}~\bibnamefont {Bass}}, \ and\
  \bibinfo {author} {\bibfnamefont {U.~W.}\ \bibnamefont {Heinz}},\ }\href
  {\doibase 10.1103/PhysRevC.89.034919} {\bibfield  {journal} {\bibinfo
  {journal} {Phys. Rev. C}\ }\textbf {\bibinfo {volume} {89}},\ \bibinfo
  {pages} {034919} (\bibinfo {year} {2014})},\ \Eprint
  {http://arxiv.org/abs/1311.0157} {arXiv:1311.0157 [nucl-th]} \BibitemShut
  {NoStop}%
\bibitem [{\citenamefont {McDonald}\ \emph {et~al.}(2017)\citenamefont
  {McDonald}, \citenamefont {Shen}, \citenamefont {Fillion-Gourdeau},
  \citenamefont {Jeon},\ and\ \citenamefont {Gale}}]{McDonald:2016vlt}%
  \BibitemOpen
  \bibfield  {author} {\bibinfo {author} {\bibfnamefont {S.}~\bibnamefont
  {McDonald}}, \bibinfo {author} {\bibfnamefont {C.}~\bibnamefont {Shen}},
  \bibinfo {author} {\bibfnamefont {F.}~\bibnamefont {Fillion-Gourdeau}},
  \bibinfo {author} {\bibfnamefont {S.}~\bibnamefont {Jeon}}, \ and\ \bibinfo
  {author} {\bibfnamefont {C.}~\bibnamefont {Gale}},\ }\href {\doibase
  10.1103/PhysRevC.95.064913} {\bibfield  {journal} {\bibinfo  {journal} {Phys.
  Rev. C}\ }\textbf {\bibinfo {volume} {95}},\ \bibinfo {pages} {064913}
  (\bibinfo {year} {2017})},\ \Eprint {http://arxiv.org/abs/1609.02958}
  {arXiv:1609.02958 [hep-ph]} \BibitemShut {NoStop}%
\bibitem [{\citenamefont {Schenke}\ \emph {et~al.}(2010)\citenamefont
  {Schenke}, \citenamefont {Jeon},\ and\ \citenamefont
  {Gale}}]{Schenke:2010nt}%
  \BibitemOpen
  \bibfield  {author} {\bibinfo {author} {\bibfnamefont {B.}~\bibnamefont
  {Schenke}}, \bibinfo {author} {\bibfnamefont {S.}~\bibnamefont {Jeon}}, \
  and\ \bibinfo {author} {\bibfnamefont {C.}~\bibnamefont {Gale}},\ }\href
  {\doibase 10.1103/PhysRevC.82.014903} {\bibfield  {journal} {\bibinfo
  {journal} {Phys. Rev. C}\ }\textbf {\bibinfo {volume} {82}},\ \bibinfo
  {pages} {014903} (\bibinfo {year} {2010})},\ \Eprint
  {http://arxiv.org/abs/1004.1408} {arXiv:1004.1408 [hep-ph]} \BibitemShut
  {NoStop}%
\bibitem [{\citenamefont {Zhu}\ \emph {et~al.}(2015)\citenamefont {Zhu},
  \citenamefont {Meng}, \citenamefont {Song},\ and\ \citenamefont
  {Liu}}]{Zhu:2015dfa}%
  \BibitemOpen
  \bibfield  {author} {\bibinfo {author} {\bibfnamefont {X.}~\bibnamefont
  {Zhu}}, \bibinfo {author} {\bibfnamefont {F.}~\bibnamefont {Meng}}, \bibinfo
  {author} {\bibfnamefont {H.}~\bibnamefont {Song}}, \ and\ \bibinfo {author}
  {\bibfnamefont {Y.-X.}\ \bibnamefont {Liu}},\ }\href {\doibase
  10.1103/PhysRevC.91.034904} {\bibfield  {journal} {\bibinfo  {journal} {Phys.
  Rev. C}\ }\textbf {\bibinfo {volume} {91}},\ \bibinfo {pages} {034904}
  (\bibinfo {year} {2015})},\ \Eprint {http://arxiv.org/abs/1501.03286}
  {arXiv:1501.03286 [nucl-th]} \BibitemShut {NoStop}%
\end{thebibliography}%


%

\appendix
\section{Values of fitting parameters}\label{fit_vars_collect}
This appendix presents values of parameters in Eqs.~(\ref{eq:fpt_levy_v0}-\ref{eq:fpt_jet}) and  (\ref{eq:fpt_levy_v2}-\ref{eq:fpt_levy_v3}) which are obtained by fitting experimental data of hadronic $p_T$ spectra in high energy collisions and are necessary for the calculation of $p_0$ and $p_1$.     

\begin{table}[h]
    \caption{Values of fitting parameters used in plotting Fig.~\ref{fig:p_c6080}. }
\centering{}%
\begin{tabular}{ccc|ccc|cc}
\toprule
\multicolumn{3}{c|}{Levy} & \multicolumn{3}{c|}{Thermal} & \multicolumn{2}{c}{Jet}\\
\midrule
$n$ & $m$ & $c$ & $v$ & $m$ & $T$ & $b$ & $n$\tabularnewline
\midrule
4.86 & 8.00 & 0.053 & 0.00 & 0.343 & 0.485 & 0.983 & 8.29\tabularnewline
\bottomrule
\end{tabular}
\end{table}

\begin{table}[h]
    \caption{Values of fitting parameters used in plotting Fig.~\ref{fig:p_c05}. }
\centering{}%
\begin{tabular}{ccc|cc}
\toprule
\multicolumn{3}{c|}{Thermal} & \multicolumn{2}{c}{Jet}\tabularnewline
\midrule
$v$ & $m$ & $T$ & $b$ & $n$\tabularnewline
\midrule
0.170 & 1.73 & 0.286 & 1.59 & 9.52\tabularnewline
\bottomrule
\end{tabular}
\end{table}

\begin{table}[h]
    \caption{Values of fitting parameters used in plotting Fig.~\ref{fig:Lam_3fit}. }
\centering{}%
\begin{tabular}{ccc|cccc|cccc}
\toprule
\multicolumn{3}{c|}{Levy} & \multicolumn{4}{c|}{Levy\_v2} & \multicolumn{4}{c}{Levy\_v3}\tabularnewline
\midrule
$n$ & $m$ & $c$ & $a$ & $n$ & $m$ & $c$ & $v$ & $n$ & $m$ & $c$\tabularnewline
\midrule
14.9 & 7.60 & 0.139 & 1.69 & 5.32 & 2.09 & 0.462 & 0.110 & 7.08 & 6.47 & 0.097\tabularnewline
\bottomrule
\end{tabular}
\end{table}


\begin{table*}
	\caption{Values of fitting parameters for different hadrons ($h$) in $pp$ collisions at $\sqrt{s}$=13 TeV, which are used in plotting Fig.~\ref{fig:pp13TeV_p0p1}.}
\centering{}%
    \begin{tabular*}{0.8\linewidth}{@{\extracolsep\fill}c|ccc|cccc|cccc}
\toprule
  & \multicolumn{3}{c|}{Levy} & \multicolumn{4}{c|}{Levy\_v2} & \multicolumn{4}{c}{Levy\_v3}\tabularnewline
\midrule
 $h$ & $n$ & $m$ & $c$ & $a$ & $n$ & $m$ & $c$ & $v$ & $n$ & $m$ & $c$\tabularnewline
\midrule
     $K_{S}^{0}$&   7.07 &  0.398 &  0.236 &  0.972 &   7.43 &  0.431 &  0.233 &  0.020 &   7.01 &  0.400 &  0.231 \\
            $K$ &   6.55 &  0.479 &  0.218 &  0.890 &   7.55 &  0.669 &  0.200 &  0.020 &   6.59 &  0.466 &  0.216 \\
       $K^{*0}$ &   9.78 &  0.542 &  0.372 &  0.856 &   14.7 &  0.799 &  0.329 &  0.300 &   9.98 &  0.412 &  0.282 \\
           $p$  &   7.69 &  0.822 &  0.264 &  0.929 &   8.16 &   1.04 &  0.237 &  0.020 &   7.81 &  0.790 &  0.262 \\
         $\phi$ &   11.6 &  0.515 &  0.435 &   1.15 &   8.22 &  0.262 &  0.500 &  0.020 &   11.6 &  0.507 &  0.427 \\
      $\Lambda$ &   16.0 &  0.577 &  0.396 &  0.898 &   22.7 &  0.800 &  0.354 &  0.020 &   16.0 &  0.568 &  0.388 \\
     $\Xi^{*0}$ &   12.3 &   1.52 &  0.414 &   1.11 &   22.7 &   1.17 &  0.491 &  0.181 &   13.3 &   1.19 &  0.364 \\
		  $\Xi$ &   17.4 &  0.884 &  0.457 &  0.891 &   23.9 &   1.25 &  0.388 &  0.020 &   17.6 &  0.866 &  0.449 \\
$\Sigma^{*\pm}$ &   10.4 &   1.53 &  0.356 &  0.899 &   11.9 &   2.00 &  0.297 &  0.020 &   10.6 &  1.492 &  0.355 \\
       $\Omega$ &   20.6 &   1.48 &  0.500 &  0.969 &   28.0 &   1.44 &  0.500 &  0.020 &   28.0 &   1.22 &  0.530 \\
\bottomrule
\end{tabular*}
\end{table*}

\begin{table*}
    \caption{Values of fitting parameters for different hadrons ($h$) in $pp$ collisions at different collision energies, which are used in plotting Fig. \ref{fig:p0p1_snn_dep_pp}.}
\centering{}%
\begin{tabular*}{0.8\linewidth}{@{\extracolsep\fill}c|c|ccc|cccc|cccc}
\toprule
 \multicolumn{2}{c|}{}  & \multicolumn{3}{c|}{Levy} & \multicolumn{4}{c|}{Levy\_v2} & \multicolumn{4}{c}{Levy\_v3}\tabularnewline
\midrule
$\sqrt{s}$ & $h$& $n$ & $m$ & $c$ & $a$ & $n$ & $m$ & $c$ & $v$ & $n$ & $m$ & $c$\tabularnewline
\midrule
   7TeV &       $K_{S}^{0}$ &   7.10 &  0.410 &  0.222 &  0.992 &   7.19 &  0.419 &  0.221 &  0.284 &  7.389 &  0.296 &  0.175 \\
   7TeV &               $K$ &   6.14 &  0.514 &  0.198 &   1.08 &   5.52 &  0.414 &  0.207 &  0.181 &   6.46 &  0.403 &  0.177 \\
   7TeV &          $K^{*0}$ &   7.70 &  0.639 &  0.297 &  0.728 &  9.199 &   1.92 &  0.180 &  0.020 &   7.76 &  0.621 &  0.293 \\
   7TeV &            $\phi$ &   6.69 &   1.08 &  0.270 &  0.796 &   5.61 &   3.57 &  0.119 &  0.020 &   6.79 &  1.036 &  0.269 \\
   7TeV &               $p$ &   10.4 &  0.635 &  0.284 &  0.927 &   12.1 &  0.771 &  0.265 &  0.058 &   10.6 &  0.594 &  0.271 \\
   7TeV &         $\Lambda$ &   13.1 &  0.752 &  0.340 &   1.08 &   10.9 &  0.600 &  0.370 &  0.064 &   13.4 &  0.697 &  0.324 \\
\bottomrule
\end{tabular*}
\end{table*}

\clearpage

\begin{longtable*}[!hp]{@{\extracolsep\fill}c|c|ccc|cccc|cccc}
    \caption{Values of fitting parameters for different hadrons ($h$) in different multiplicity classes in $pp$ collisions at $\sqrt{s}$= 13 TeV, which are used in plotting Fig.~\ref{fig:pp13TeV_p0p1_nch_dep}.}\\
\toprule
 \multicolumn{2}{c|}{}  & \multicolumn{3}{c|}{Levy} & \multicolumn{4}{c|}{Levy\_v2} & \multicolumn{4}{c}{Levy\_v3}\tabularnewline
\midrule
$h$ & class & $n$ & $m$ & $c$ & $a$ & $n$ & $m$ & $c$ & $v$ & $n$ & $m$ & $c$\tabularnewline
\midrule
\endfirsthead
\midrule
$h$ & class & $n$ & $m$ & $c$ & $a$ & $n$ & $m$ & $c$ & $v$ & $n$ & $m$ & $c$\tabularnewline
\midrule
\endhead
            $K$ &           I &   7.63 &  0.395 &  0.302 &   1.09 &   1.09 &  0.300 &  0.317 &  0.020 &   7.54 &  0.400 &  0.295 \\
            $K$ &          II &   7.24 &  0.484 &  0.275 &   1.04 &   6.75 &  0.430 &  0.282 &  0.174 &   7.39 &  0.400 &  0.238 \\
            $K$ &         III &   6.99 &  0.492 &  0.257 &   1.07 &   6.29 &  0.409 &  0.268 &  0.193 &   7.13 &  0.400 &  0.219 \\
            $K$ &          IV &   6.82 &  0.495 &  0.245 &   1.07 &   6.13 &  0.409 &  0.255 &  0.192 &   6.97 &  0.400 &  0.210 \\
            $K$ &           V &   6.79 &  0.476 &  0.239 &   1.05 &   6.28 &  0.415 &  0.245 &  0.169 &   6.89 &  0.400 &  0.208 \\
            $K$ &          VI &   6.67 &  0.479 &  0.227 &   1.06 &   6.12 &  0.411 &  0.234 &  0.172 &   6.77 &  0.400 &  0.198 \\
            $K$ &         VII &   6.54 &  0.484 &  0.213 &   1.07 &   5.94 &  0.405 &  0.220 &  0.173 &   6.66 &  0.400 &  0.186 \\
            $K$ &        VIII &   6.44 &  0.497 &  0.199 &   1.08 &   5.72 &  0.398 &  0.208 &  0.197 &   6.56 &  0.400 &  0.171 \\
\bottomrule
\end{longtable*}


\begin{longtable*}{@{\extracolsep\fill}c|c|ccc|cccc|cccc}
    \caption{Values of fitting parameters for different hadrons ($h$) in different multiplicity classes in $pp$ collisions at $\sqrt{s}$= 7 TeV, which are used in plotting Fig.~\ref{fig:pp7TeV_p0p1_nch_dep}.}\\
\toprule
 \multicolumn{2}{c|}{}  & \multicolumn{3}{c|}{Levy} & \multicolumn{4}{c|}{Levy\_v2} & \multicolumn{4}{c}{Levy\_v3}\tabularnewline
\midrule
$h$ & class & $n$ & $m$ & $c$ & $a$ & $n$ & $m$ & $c$ & $v$ & $n$ & $m$ & $c$\tabularnewline
\midrule
\endfirsthead
\midrule
$h$ & class & $n$ & $m$ & $c$ & $a$ & $n$ & $m$ & $c$ & $v$ & $n$ & $m$ & $c$\tabularnewline
\midrule
\endhead
    $K_{S}^{0}$ &           I &   8.28 &  0.442 &  0.293 &  0.959 &   8.96 &  0.500 &  0.285 &  0.022 &   8.30 &  0.432 &  0.288 \\
    $K_{S}^{0}$ &          II &   8.08 &  0.423 &  0.276 &  0.945 &   8.96 &  0.500 &  0.267 &  0.075 &   8.14 &  0.392 &  0.259 \\
    $K_{S}^{0}$ &         III &   7.78 &  0.424 &  0.261 &  0.945 &   8.60 &  0.500 &  0.252 &  0.076 &   7.84 &  0.392 &  0.244 \\
    $K_{S}^{0}$ &          IV &   7.63 &  0.422 &  0.250 &  0.943 &   8.43 &  0.500 &  0.241 &  0.135 &   7.73 &  0.366 &  0.223 \\
    $K_{S}^{0}$ &           V &   7.46 &  0.425 &  0.240 &  0.945 &   8.19 &  0.500 &  0.232 &  0.031 &   7.48 &  0.412 &  0.234 \\
    $K_{S}^{0}$ &          VI &   7.47 &  0.412 &  0.233 &  0.971 &   7.84 &  0.447 &  0.229 &  0.022 &   7.64 &  0.324 &  0.193 \\
    $K_{S}^{0}$ &         VII &   7.18 &  0.428 &  0.217 &  0.976 &   7.47 &  0.457 &  0.215 &  0.122 &   7.33 &  0.372 &  0.198 \\
    $K_{S}^{0}$ &        VIII &   6.90 &  0.447 &  0.202 &   1.03 &   6.55 &  0.407 &  0.206 &  0.021 &   7.16 &  0.345 &  0.172 \\
\bottomrule
\end{longtable*}

\clearpage

\begin{longtable*}{@{\extracolsep\fill}c|c|ccc|cccc|cccc}
    \caption{Values of fitting parameters for different hadrons ($h$) in different multiplicity classes (V0A percentile) in $p$-Pb collisions at $\sqrt{s_{NN}}$=5.02 TeV, which are used in plotting Fig.~\ref{fig:pPb5TeV_p0p1_nch_dep}. }\\
\toprule
 \multicolumn{2}{c|}{}  & \multicolumn{3}{c|}{Levy} & \multicolumn{4}{c|}{Levy\_v2} & \multicolumn{4}{c}{Levy\_v3}\tabularnewline
\midrule
$h$ & $pct$ & $n$ & $m$ & $c$ & $a$ & $n$ & $m$ & $c$ & $v$ & $n$ & $m$ & $c$\tabularnewline
\midrule
\endfirsthead
\midrule
$h$ & $pct$ & $n$ & $m$ & $c$ & $a$ & $n$ & $m$ & $c$ & $v$ & $n$ & $m$ & $c$\tabularnewline
\midrule
\endhead
            $K$ &       0-5\% &   11.0 &  0.475 &  0.324 &  0.873 &   15.5 &  0.699 &  0.291 &  0.020 &   11.0 &  0.470 &  0.321 \\
            $K$ &      5-10\% &   11.3 &  0.426 &  0.324 &  0.912 &   14.6 &  0.555 &  0.305 &  0.030 &   11.3 &  0.414 &  0.316 \\
            $K$ &     10-20\% &   10.4 &  0.432 &  0.310 &  0.848 &   15.6 &  0.692 &  0.276 &  0.210 &   10.9 &  0.344 &  0.259 \\
            $K$ &     20-40\% &   9.51 &  0.415 &  0.294 &  0.865 &   13.3 &  0.622 &  0.268 &  0.209 &   9.85 &  0.332 &  0.245 \\
            $K$ &     40-60\% &   8.87 &  0.400 &  0.272 &  0.853 &   12.8 &  0.603 &  0.250 &  0.010 &   9.06 &  0.384 &  0.272 \\
            $K$ &     60-80\% &   7.64 &  0.409 &  0.236 &  0.929 &   8.69 &  0.500 &  0.228 &  0.084 &   7.73 &  0.374 &  0.221 \\
            $K$ &    80-100\% &   6.80 &  0.461 &  0.189 &  0.972 &   7.05 &  0.500 &  0.185 &  0.182 &   7.05 &  0.368 &  0.166 \\
\midrule
    $K_{S}^{0}$ &       0-5\% &   10.5 &  0.492 &  0.319 &  0.995 &   10.6 &  0.500 &  0.317 &  0.060 &   10.7 &  0.459 &  0.303 \\
\bottomrule
\end{longtable*}

\clearpage
\begin{longtable*}{@{\extracolsep\fill}c|c|c|ccc|cccc|cccc}
    \caption{Values of fitting parameters for different hadrons ($h$) in different collision centralities ($pct$) in/at different collision systems/energies ($colls$), which are used in plotting Fig.~\ref{fig:p0_AA_npart}. }\\
\toprule
 \multicolumn{3}{c|}{}  & \multicolumn{3}{c|}{Levy} & \multicolumn{4}{c|}{Levy\_v2} & \multicolumn{4}{c}{Levy\_v3}\tabularnewline
\midrule
$colls$ & $h$ & $pct$ & $n$ & $m$ & $c$ & $a$ & $n$ & $m$ & $c$ & $v$ & $n$ & $m$ & $c$\tabularnewline
\midrule
\endfirsthead
\midrule
$colls$ & $h$ & $pct$ & $n$ & $m$ & $c$ & $a$ & $n$ & $m$ & $c$ & $v$ & $n$ & $m$ & $c$\tabularnewline
\midrule
\endhead
\multirow{ 28 }{*}{  }
                 &             $K$ &      0-5\% &   14.0 &  0.957 &  0.282 &   1.39 &   7.90 &  0.300 &  0.431 &  0.020 &   14.4 &  0.916 &  0.280 \\
               ~ &             $K$ &     5-10\% &   10.8 &   1.31 &  0.236 &  0.835 &   8.11 &   4.13 &  0.102 &  0.020 &   11.0 &   1.24 &  0.236 \\
               ~ &             $K$ &    10-20\% &   11.9 &  0.974 &  0.270 &  0.835 &   8.32 &   3.21 &  0.123 &  0.020 &   12.2 &  0.930 &  0.269 \\
               ~ &             $K$ &    20-30\% &   11.0 &  0.921 &  0.265 &  0.848 &   5.90 &   5.64 &  0.069 &  0.020 &   12.3 &  0.882 &  0.264 \\
               ~ &             $K$ &    30-40\% &   9.58 &  0.903 &  0.251 &  0.913 &   10.1 &   1.24 &  0.215 &  0.020 &   9.73 &  0.867 &  0.250 \\
               ~ &             $K$ &    40-50\% &   8.79 &  0.823 &  0.243 &  0.888 &   9.57 &   1.22 &  0.202 &  0.020 &   8.90 &  0.791 &  0.241 \\
               ~ &             $K$ &    50-60\% &   8.50 &  0.711 &  0.238 &  0.877 &   9.85 &   1.05 &  0.204 &  0.020 &   8.57 &  0.689 &  0.236 \\
               ~ &             $K$ &    60-70\% &   7.55 &  0.652 &  0.225 &   1.10 &   6.62 &  0.505 &  0.242 &  0.060 &   7.68 &  0.602 &  0.216 \\
\bottomrule
\end{longtable*}

\end{document}